\documentclass[aps,pra,twocolumn,superscriptaddress,showpacs,preprintnumbers,amsmath,amssymb,longbibliography]{revtex4-1}
\usepackage{epsfig}
\usepackage{natbib}
\usepackage{amssymb}
\usepackage{soul}
\usepackage[breaklinks=true,colorlinks]{hyperref}
\newcommand{\bn}{\begin{eqnarray}}
\newcommand{\en}{\end{eqnarray}}
\newcommand{\eml}{\end{multline}}
\newcommand{\bml}{\begin{multline}}

\begin{document}
 
\title{Work Generation from Thermal Noise by Quantum Phase-Sensitive Observation}
\author{Tomas Opatrn\'y}
\affiliation{Department of Optics, Faculty of Science, Palack\'y University, 17. listopadu 50, 77146 Olomouc, Czech Republic}
\author{Avijit Misra}
\email{avijitmisra0120@gmail.com}
\affiliation{International Center of Quantum Artificial Intelligence for Science and Technology (QuArtist)
and Department of Physics, Shanghai University, 200444 Shanghai, China}
\affiliation{Department of Chemical and Biological Physics,
Weizmann Institute of Science, Rehovot 7610001, Israel}

\author{Gershon Kurizki}
\affiliation{Department of Chemical and Biological Physics,
Weizmann Institute of Science, Rehovot 7610001, Israel}

\begin{abstract}
We put forward the concept of work extraction from thermal noise by phase-sensitive (homodyne) measurements of the noisy input followed by (outcome-dependent) unitary manipulations of the post-measured state. For optimized measurements, noise input with more than one quantum on average is shown to yield heat-to-work conversion with efficiency and power that grow with the mean number of input quanta, detector efficiency and its inverse temperature. This protocol is shown to be advantageous compared to common models of information and heat engines.
\end{abstract}

\date{\today}
\maketitle
 
\begin{figure*}
  \centering
  \includegraphics[width=12cm]{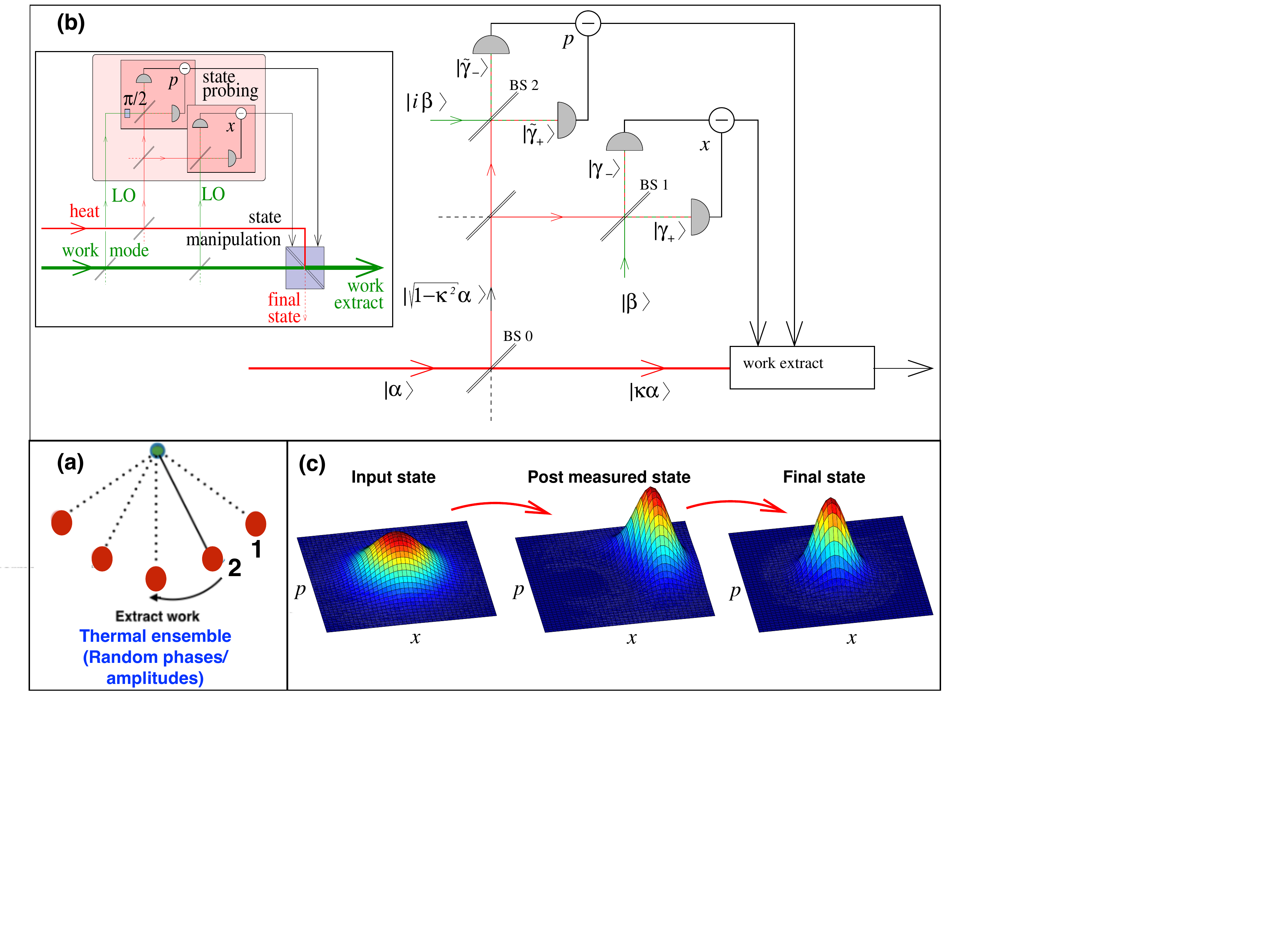}
  \caption{(a)Work extraction by snapshots (1,2) from a random ensemble of pendula. (b) WOF scheme: for a thermal mixture of coherent states $|\alpha \rangle$). A homodyne measurement (see text)  is performed  on the reflected, weak part of the input superposed with a (comparably weak) local oscillator (LO) to optimally estimate the quadratures $x$ and $p$. The result is used to adjust the output to constructively interfere with the LO and thereby downshift it to extract work. (c) A thermal mixture of coherent states is transformed  by the measurement to a displaced, squeezed (slightly non-Gaussian) state. Work is extracted by displacement and unsqueezing to a state with much less energy than the post-measured state.}
  \label{Fig1}
\end{figure*}

{\it Introduction.}-- The highest entropy at a given energy pertains to thermal noise, which is a ubiquitous form of energy in the universe \cite{Gardiner}. Since work \cite{Alicki_1979,Hanggi2007} is an {\it ordered} form of energy, delivered {\it without entropy change} \cite{Pusz_1978,Gelbwaser_2013_b,PRE2014,David15,niedenzu18quantum}, a thermal ensemble of oscillators stores heat but not work. Here we propose an efficient way to harness such ensembles for fast performance of useful work.
Classically, the protocol appears to be straightforward: impulsively observe the phase and amplitude of each oscillator (via two ``snapshots'' at a chosen time interval), wait until it is in full swing, then let it discharge its stored work (Fig. \ref{Fig1}a).
Yet, what is the quantum mechanical (QM) counterpart of this protocol? 
Any noisy ensemble of QM harmonic oscillators at a given frequency (mode) forms a random distribution of coherent states. Therefore, our QM protocol invokes homodyne measurements \cite{Gardiner,ScullyBOOK97,Paris,Schleichbook01,Scully_BOOK,Carmichael_BOOK,Walls_BOOK,Ulf,ParisHomod} optimized to approximately reveal a coherent-state component of the random distribution, and thereby sample the quadratures of the oscillator field within the uncertainty-limit accuracy. 
We show that unitary manipulations of the post-measured state that are determined by the measurement outcome can yield heat-to-work conversion at an  efficiency that grows with input temperature.

 This protocol introduces the concept of exploiting {\it randomly distributed,  non-commuting, continuous variables} as thermodynamic resources for work extraction by estimating their quadrature values {\it at minimal energy cost}. We dub it {\it work by observation and feedforward} (WOF).


{\it WOF engine principles.}--   We consider an input state of a harmonic oscillator, e.g.-- a single electromagnetic field-mode, whose phase-space distribution 
falls off monotonically and isotropically from its zero-energy (vacuum) origin \cite{Schleichbook01,ScullyBOOK97}, as in the case of the Gaussian thermal state.
Such a QM state, dubbed {\it passive} \cite{Pusz_1978}, is incapable of delivering work by unitary transformations.
It must be rendered non-passive to allow for subsequent work extraction from its stored work (alias ergotropy) by a unitary process \cite{Allahverdyan_2000,Pas4,Pas6,Gelbwaser_2015_b,Gelbwaser_2013_b,David15,PRE2014,Ghosh2017,Ghosh2018} (SM-1). A standard homodyne measurement  can transform this passive state into a non-passive coherent state by mixing it with a much stronger, coherent, local oscillator (LO) \cite{ScullyBOOK97,Paris,Schleichbook01,Scully_BOOK,Carmichael_BOOK,Walls_BOOK}.
Yet, 
to extract maximal work, the measurement should consume as little energy as possible. How can this be achieved?

\begin{figure}
 \hspace{-0.5cm} \includegraphics[width=8.7cm]{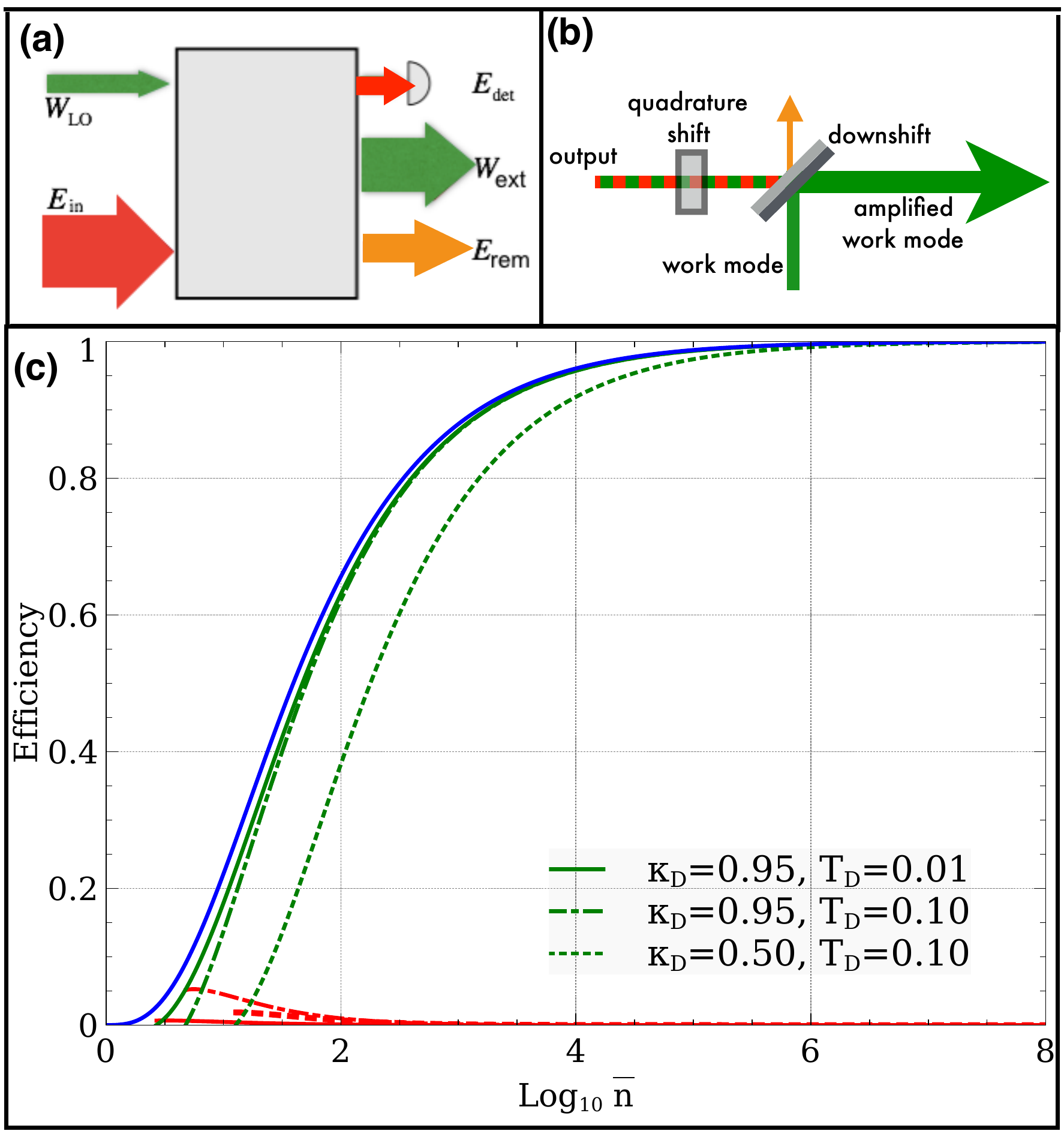}
  \caption{(a) Energy balance of the WOF scheme. The passive input state has mean energy $E_{\rm in} = \hbar \omega\bar n$, the local oscillators (LO) have mean energy $E_{\rm LO} = 2\hbar \omega\beta^2$. The detectors absorb energy $E_{\rm det}$. $W$ is extracted by displacement and unsqueezing of the unmeasured main fraction of the input.  The remaining energy $E_{\rm rem}$ (orange) is unexploitable as work. Work change--green, heat exchange--red. (b) Work extraction by BS transmittance and phase-shift changes causing constructive interference of the output with the coherent working mode. Orange arrow--remaining (typically thermal) passive output. (c) WOF efficiency as function of ${\rm Log}_{10} \bar n$ (Eq. (\ref{effi})): the bound $\eta^{(1)}_{max}
  $ (blue) and actual $\eta$ for different $\kappa_D$ and scaled detector temperatures $T_D\leftrightarrow k_BT_D/\hbar \omega$ (green solid, dashed dot, dotted and dashed). The red lines show $Q_{\rm reset}/E_{\rm in}$ for the same parameter values as their green counterparts.
  Depending on $\kappa_D$, the impact of the resetting cost $Q_{\rm reset}$ on the efficiency is seen to be negligible for sufficiently large $\bar n$. In these plots thermal noise in the local oscillator (LO) and the detectors have mean photon numbers
$\bar n_{\rm LO} = \bar n_{\rm D} = 0.05$ (see SM-5).
}
  \label{Fig2}
\end{figure}

To this end, we propose a non-standard homodyne measurement that only probes a split-off small fraction of the thermal input field
by mixing it with  a  LO as weak as this fraction (Fig.~\ref{Fig1}b). This measurement yields quadrature values of the field with optimal tradeoff between energy cost and precision. The measurement outcome serves to determine the unitary operations that extract maximal work from the post-measured output: 
%
%
a {\it downshift} (displacement) towards the zero-energy origin, supplemented by {\it unsqueezing} (Fig.~\ref{Fig1}c).
The downshift can be realized by adjusting the transmissivity and phase delay of a beam splitter (or an amplitude-phase modulator) according to the outcome. The output-field quadratures are then shifted by this beam splitter to make the output constructively interfere with the coherent field in the working mode (Fig. \ref{Fig2} a,b). For $\bar n \gg 1$, $\bar n$ being the mean number of input quanta, nearly the {\it entire energy} of the thermal ensemble is shown to be extractable as work, with efficiency $1-O(1/\sqrt{\bar n})$, {\it by a single optimized homodyne measurement}. The energy cost that may limit the WOF efficiency is accounted for, the fundamental cost being the detector-record erasure (resetting) cost\cite{LandauerIBM61,Berut_2012,lutz2015information,goold2014nonequilibrium,
QuantitativeLandauer,AlickiARXIV10}.
%

The WOF scheme is feasible and conceptually simple (Fig.~\ref{Fig1}c, Fig.~\ref{Fig2} a,b). It is shown to be advantageous compared to Szilard/ Maxwell-Demon information engines based on binary measurements of discrete variables \cite{Maxwell,Szilard1929,SagawaPRL08,Kim_2011,Park_2013,Diaz,parrondo2015thermodynamics, R2, Vid2016,Strunz2019,Bengtsson2018,Chida2017,Aydin2020}. It can also outperform common models of heat engines that exploit the same resources (see Discussion).

{\it Work extraction and its bounds.}-- Any single-mode input state can be represented as:
$
\hat \varrho = \int \!  \int P(\alpha) |\alpha \rangle \langle \alpha | d^2 \alpha,
$
$P$ being the Glauber-Sudarshan distribution function of coherent-states $|\alpha\rangle$ with complex amplitudes $\alpha$ \cite{Schleichbook01,Scully_BOOK,Carmichael_BOOK,Walls_BOOK}.
 Let us first consider a coherent-state component $|\alpha\rangle$ of the input distribution (Fig.~\ref{Fig1}b). After the $0$th beam splitter (BS0) with high transmissivity $\kappa$,
 the state $|\kappa \alpha\rangle$ is transmitted and the state $|\sqrt{1-\kappa^2}\alpha \rangle$ (that has a much smaller amplitude) is reflected (split off) towards the homodyne detectors. These detectors serve for estimating the quadrature operators $\hat x$ and $\hat p$, $\hat x = 2^{-1/2}(\hat a + \hat a^{\dag})$, $\hat p = -2^{-1/2}i(\hat a - \hat a^{\dag})$, where $[ \hat a, \hat a^{\dag}]=1$ (here we set $\hbar = \omega =1$).
To effect the estimations, the  small split-off fractions are superposed at the detectors with 
 two LOs at the same frequency $\omega$. The two LOs (originating from a common source) are prepared by two BS and a $\frac{\pi}{2}$ phase shifter in coherent states $| \beta\rangle$  and $| i\beta\rangle$ with orthogonal quadrature amplitudes, $\beta$ being chosen real (Fig.~\ref{Fig1}b).
Behind BS1 and BS2 we then have a $4$-mode coherent state
$|\psi \rangle = |\gamma_+\rangle|\gamma_-\rangle|\tilde\gamma_+\rangle|\tilde\gamma_-\rangle$ with amplitudes $\gamma_{\pm} =\frac{1}{\sqrt{2}}\left(\sqrt{\frac{1-\kappa^2}{2}}\alpha \pm \beta  \right),$ $\tilde\gamma_{\pm} =\frac{1}{\sqrt{2}}\left(\sqrt{\frac{1-\kappa^2}{2}}\alpha \pm i\beta  \right).$
%
Photodetection of states $|\gamma_{\pm}\rangle$, $|\tilde\gamma_{\pm}\rangle$ yields Poissonian statistics with mean values $\bar n_{\pm} = |\gamma_{\pm}|^2$, $\bar{\tilde{n}}_{\pm} = |\tilde\gamma_{\pm}|^2$.
A homodyne measurement \cite{Schleichbook01,Ulf,ScullyBOOK97,Scully_BOOK,Carmichael_BOOK,Walls_BOOK,ParisHomod} consists in recording  photocount differences between the two pairs of detectors, $\Delta n_x \equiv n_+-n_-$ and  $\Delta n_p \equiv \tilde {n}_+-\tilde{n}_-$. These $\Delta n_x$, $\Delta n_p$ carry information on the quadrature eigenvalues $x$ and $p$,  since $\bar n_{\pm}$ and  $\bar{\tilde{n}}_{\pm}$ and their variances depend on $\alpha=\frac{1}{\sqrt{2}}(x+ip)$ (SM-2). 
 
 The probability distribution for $\Delta n_{x}$ and $\Delta n_{p}$ on condition that the input state was $|\alpha\rangle$, $P(\Delta n_{x},\Delta n_{p}|\alpha) = P(\Delta n_{x}|\alpha)P(\Delta n_{p}|\alpha)$, 
can be inverted
by means of the Bayes rule.
The post-measurement state conditional on $\Delta n_{x}$ and $\Delta n_{p}$ that characterizes the unmeasured (transmitted) part of the output for {\it any distribution} $ P(\alpha)$ has then the form
\begin{eqnarray}
\hat \varrho (\Delta n_{x},\Delta n_{p})= \frac{1}{\kappa^2} \int \! \int P\left(\frac{\alpha}{\kappa}|\Delta n_{x},\Delta n_{p}\right) |\alpha \rangle \langle \alpha | d^2 \alpha
\end{eqnarray}
We start from a thermal state with Gaussian $P(\alpha)$, but the resulting state is in general a nonpassive state (unless $\Delta n_{x}=\Delta n_{p}= 0$ ) (Fig. \ref{Fig1}c).

The measured $\Delta n_{x}$, $\Delta n_{p}$ determine the required downshift (displacement) of the output state towards a state whose mean quadratures are zero. This yields work extraction in the amount $W(\Delta n_{x}, \Delta n_{p})$ (SM-2).
The mean work obtained following such displacement, but ignoring the resetting cost of the detectors (considered below), can be found by averaging $W(\Delta n_{x}, \Delta n_{p})$ over the probability distribution $P(\Delta n_{x},\Delta n_{p})$ and subtracting the invested  mean energy of the two orthogonal-quadrature LOs, $2\hbar\omega\beta^2$, to yield the mean work
\begin{eqnarray}
W = \sum_{\Delta n_{x}} \sum_{\Delta n_{x}}W (\Delta n_{x}, \Delta n_{p})
P(\Delta n_{x},\Delta n_{p}) - 2\hbar\omega\beta^2 .
\label{EWnet}
\end{eqnarray} 
%
%
%
Under the Gaussian approximation (SM-2), one can analytically maximize this extractable mean work with respect to the BS0 transmissivity  $\kappa$ and the intensity $\beta^2$ of the LO. This maximization of the mean work in Eq. (\ref{EWnet}) yields (SM-3)
\begin{eqnarray}
W_{\rm max} \approx
\hbar \omega\left(\sqrt{\bar n - \sqrt{\bar n}+1}-1 \right)^2  \left( 1-\frac{1}{\sqrt{\bar n}} \right).
\label{eq3}
\end{eqnarray}
Equation (\ref{eq3}) indicates that mean work extraction ($W_{\rm max}>0$) by WOF requires thermal input with $\bar n>1$.
For $\bar{n}\gg1$ the optimal LO $\beta^2\sim\sqrt{\bar{n}}$ is {\it much weaker} than the {\it input} signal, the opposite of standard homodyning \cite{Schleichbook01,ScullyBOOK97}.


 A  displacement transformation that maximally downshifts  the post-measured state in energy does not fully extract the work from it, since the downshifted state is in general not passive, and still keeps work capacity (ergotropy, SM-1). To extract more work, we can apply an {\it unsqueezing transformation} (by a Kerr medium \cite{Scully_BOOK,Carmichael_BOOK,Walls_BOOK}) to the downshifted state that is centered at the origin, with $\langle x \rangle=\langle p \rangle=0$. This state has a 
  mean energy of $E_0=\frac{\hbar \omega}{2}( \langle \hat x^2 \rangle +\langle \hat p^2 \rangle ) = \frac{\hbar \omega}{2}(V_++V_-)$, where $V_{\pm}$ are the eigenvalues of the variance matrix of  $\hat x$ and $\hat p$ (SM-4).
The minimal energy state attainable by unsqueezing has the energy $E_{\rm min}=\hbar \omega\sqrt{V_+V_-}$ with $V_+=V_-$. Upon averaging the work extractable  by unsqueezing, $W_{US} (\Delta n_{x},\Delta n_{p}) = E_0-E_{\rm min}$,
over all  measured values of $\Delta n_{x}$, $\Delta n_{x}$, we  find that the work in Eq. (2) increases on account of $W_{US}$ by $18\%$ for $\bar n=2$, $12\%$ for $\bar n=5$ and so on: {\it $W_{US}$ only matters for small $\bar n$} (SM-4).

Hence, {\it at high temperature} ($\bar n \gg 1$), the maximal work extraction from an input with mean energy $E_{\rm in}=\hbar \omega \bar n$ coincides with the work by displacement in Eq. (\ref{eq3}) which reduces to
\begin{eqnarray}
W_{\rm max} \approx \hbar \omega\left[\bar n - 4\sqrt{\bar n} + 6  +O\left(\frac{1}{\sqrt{\bar n}} , \frac{1}{\bar n}\right)\right].
\label{EWoptHot}
\end{eqnarray}
The $4\hbar \omega\sqrt{\bar n}$ cost is the sum of the LO energy $E_{\rm LO}\approx\hbar \omega\sqrt{\bar n}$, the input energy fractions absorbed by the detectors $E_{\rm abs}\approx\hbar \omega\sqrt{\bar n}$ and the remaining (unexploited) output energy $E_{\rm rem}\approx2\hbar \omega\sqrt{\bar n}$.
This $E_{\rm rem}$ corresponds to the (typically thermal) output fluctuations and reflects the fact that our approximate measurement prepares a mixed state that cannot be unitarily transformed to the vacuum state.

The process outlined above can be iterated to exploit $E_{\rm rem}$ for more work extraction and higher efficiency, taking at the $k-$th step $E_{\rm rem}^{(k)}=\hbar \omega\bar{n}_k=2\hbar \omega\sqrt{ \bar{n}_{k-1}}$ for $k=1,...,N.$
We should stop the $N$ iterations for  $\bar n_k$ just barely above 1, at which point only negligible work is added, $W_{\rm (max)}^{(k)} \approx \hbar \omega(\bar n_k -1)^3/32.$ Practically, these iterations do not significantly increase the work output (SM-3).



To sustain WOF operation, we must reset the detectors after each work-extraction step. The energy cost of such resetting
\cite{lutz2015information,AlickiARXIV10,Berut_2012,goold2014nonequilibrium,QuantitativeLandauer,LandauerIBM61}, $Q_{\rm reset}$, sets the fundamental threshold of WOF to be $W_{\rm net}=W-Q_{\rm reset}>0$. Detector resetting to the initial temperature $T_D$ requires  a minimal energy $Q_{\rm reset}=Ik_BT_D\ln2$, where $I$ is the mean information stored (in bits) by the detectors (SM-6). For $\bar n \gg1$, $I\simeq \frac{1}{2}\ln (\bar n /4)$. Since only a small fraction of the signal is detected ($\Delta \bar n_d$ quanta in SM-6), $Q_{\rm reset}$ is negligible compared to the mean input energy $E_{\rm in}=k_BT$ when $\bar n\gg 1$, $T\gg T_D$. The resetting cost scales much slower (in orders of magnitude) with $\bar n$ than the work (Fig. \ref{Fig2}c and Fig. \ref{F2S} in SM-6)


The WOF efficiency, defined as the ratio of the net work output to the heat input, is bounded after the first measurement by
\begin{eqnarray}
\label{effi}
 \eta= \frac{W_{\rm net}}{E_{\rm in}}<\eta_{\rm max}^{(1)}= \frac{W_{\rm max}}{E_{\rm in}}.
\end{eqnarray}



Eq. (\ref{effi}) refers to the fundamental (``internal'') WOF efficiency $\eta$. The heat-to-work conversion threshold is $\eta>0$. Imperfect photodetector efficiency ($\kappa_D^2<1$) and finite temperature ($T_D>0$) obviously raise this threshold (SM-5). As seen from Fig. (\ref{Fig2}c) (Fig. \ref{F2S} in SM-6), the WOF threshold and efficiency are close to the maximal bound in Eq. (\ref{effi}) for existing highly-efficient and cold photodetectors \cite{Natarajan, Wolff2020}.

{\it Discussion.}--
We have introduced a simple scheme for WOF -- the hitherto unexplored heat-to-work conversion via information acquisition on
{\it continuous variables} of random (quantum or classical) single-mode fields. The WOF scheme adheres to the laws of thermodynamics: part of the thermal input energy is transferred to the working  mode with much less entropy than the input mode, the rest of the entropy is distributed between the detectors and the unexploited (remaining) output. WOF can be thought of as an information-based maser/laser (IBM): an amplifier of coherent signals at the expense of information that allows extracting the quadrature values of a thermal pump (input). Its efficiency is defined analogously to that of a laser or maser \cite{Scully_BOOK}, as the ratio of the output (signal) to the input (pump) energy. 

At the heart of WOF is the ability to estimate the quadratures at minimal energy cost: 
Unlike standard homodyning \cite{Ulf,Schleichbook01,ScullyBOOK97,ParisHomod}, the local oscillator (LO) and the measured field are chosen to be {\it small fractions} $\sim \sqrt{\bar n}$ of the mean input $\bar n$, optimizing the work-information tradeoff.
In order to extract maximal power and work (within the bounds of Eq. (\ref{effi})), the WOF protocol duration must only exceed the resetting time $\tau_{\rm reset}$ of the detectors to their initial temperature $T_D$ (SM-6)  \cite{Gaudenzi2018}. In existing 
photodetectors \cite{Natarajan,Wolff2020} $\tau_{\rm reset}\gtrsim 10$ nsec at the cost of $\sim 10$ times the detected photon energy $\hbar\omega$.
To boost the power, $\tau_{\rm reset}$ can be made much shorter than the natural relaxation time of the excited detector level (or band) by resetting in the non-Markovian anti-Zeno regime \cite{Erez2008,Gordon_2009,prl10,Mukherjee2020} at a modest energy cost $\sim\hbar/t_C\ll\hbar \omega$, where $t_C$ is the correlation (memory) time of the environment.


Although their principle of operation is completely different, it is instructive to compare the performance of WOF and heat engines (HE) with similar resources. For this, let us assume that both engines are energized by a hot bath with the energy $E_{\rm in}=k_BT_h$ and the HE cold bath is chosen to have the energy $k_BT_c=E_{\rm rem}$ (Eq. (\ref{effi})) (although $E_{\rm rem}$ may not be associated with a genuine cold bath). 
By this choice, the idealized HE Carnot bound at the reversibility point, $\eta_{\rm Carnot}=1-T_c/T_h$ is {\it formally} equated to the {\it hypothetical} efficiency bound of WOF had it been reversible, i.e., free of measurement costs, $\eta_{\rm reverse}\equiv1-E_{\rm rem}/E_{\rm in}$.
Yet, even with this choice, HE and WOF can perform very differently: HE power production vanishes at the Carnot bound, and the efficiency bound at the maximal work point of generic HE can be much lower \cite{Rezek,Rezek_2006,Gelbwaser_2015,Kosloff_2013,FTE5,raja2020finitetime} (SM-8), whereas $\eta_{\rm reverse}$ is similar to the bound $\eta_{\rm max}^{(1)}$ of the WOF that corresponds to maximal work production $\eta_{\rm reverse}\simeq\eta_{\rm max}^{(1)}\simeq \eta \simeq 1-O(1/\sqrt{\bar n})$ for $\bar n \gg1$ (Fig.~\ref{Fig2}c). In general, there is an inherent (model-dependent) tradeoff between HE power and efficiency \cite{Rezek,Rezek_2006,Gelbwaser_2015,Kosloff_2013,FTE5,raja2020finitetime}, since the work and power production are reduced at excessively short cycles due to friction or incomplete heat exchange with the heat baths \cite{Rezek,Rezek_2006}. By contrast, there is no such tradeoff in WOF, where power grows with the process rate provided it is less than $1/\tau_{\rm reset}$. Therefore, WOF may in principle outperform common HE with same resources, e.g., the Otto HE, (SM-8). A fully quantitative comparison of HE and WOF is unfeasible since the efficiency and power output of realistic HE are generally lower than the theoretical bounds \cite{Rezek,Rezek_2006}, partly due to on- and off- switching of their coupling to heat baths and controlling the adiabatic steps \cite{delcampo, ozguer,OttoRonnie} whose energy cost must be accounted for. Likewise, WOF feedforward cost cannot be simply estimated (see below).

 For a given $E_{\rm in}=k_BT\gg\hbar \omega (\bar n\gg1)$, the upper bounds on work production efficiency  in our WOF scheme may well surpass those of a Szilard/ Maxwell-Demon binary decision engine energized by thermal-noise photodetection \cite{Vid2016}, since WOF consumes only a $O(1/\sqrt{\bar n})$ fraction of the input, whereas its Szilard counterpart consumes a fraction comparable to 1 (SM-7). 
 For $T\rightarrow \infty$ (the classical limit) WOF is at its best, since homodyning then does not require photon counting, but merely snapshots with negligible energy cost: For example a thermal ensemble of classical pendula with mean energy of 1 erg and frequency of 1 Hz contains $\bar n\sim 10^{27}$ (which need not be counted, only the pendula motion needs to be photographed for WOF), WOF then has $\sim1-10^{-13}$ efficiency, which can hardly be surpassed by other methods!
With currently available detector efficiency $\kappa_D^2\gtrsim0.9$ and temperature $T_D\lesssim 1 mK$ \cite{Natarajan,Wolff2020}, only a few photons, $\bar n\lesssim 10$,
suffice to generate work output, i.e. a much less noisy signal than the input (Fig. \ref{Fig2}c). The hard lower bound  on  $W_{\rm net}$ production is the Landauer resetting bound (SM-6). Since the reset energy cost is currently ca. $10$- fold \cite{Natarajan,Wolff2020}, $\bar n \gtrsim 10^2$ practically ensure $\eta >0$ in Eq. (\ref{effi}).

By definition, {\it all information engines}, including WOF, have technical energy costs of signal processing and the conversion of this information into physical manipulations required for feedforward, but these costs are commonly disregarded \cite{SagawaPRL08,Kim_2011,Park_2013,Diaz,parrondo2015thermodynamics, R2, Vid2016,Strunz2019,Bengtsson2018,Chida2017,Aydin2020,Szilard1929,Maxwell,Elouard18,Elouard17}. One can treat such technical costs  as extra energy consumption that sets the threshold for {\it autonomous} WOF. Yet, these thresholds are strongly setup-dependent and therefore cannot be generally quantified. Thus, standard photodetection and electro-optical feedforward techniques can be replaced by all-optical techniques that may demand much smaller energy: 1) {\it quantum-nondemolition photon counting} of the signal by an optical probe in Rydberg polaritonic media \cite{Friedler,Shahmoon2011,Gorshkov,Firstenberg2013,Tiarks2019}; 2) output signal processing by unconventional {\it heat-powered} transistors \cite{PhysRevLett.116.200601,PhysRevE.99.042102,Tahir}; and 3) photorefractive beam splitters that can control the output quadrature shifts by signal-pump interference \cite{boyd}.

The WOF scheme is generally applicable to any noisy source (not only thermal), where homodyning of continuous variables can be performed, e.g., in ultracold bosonic gases where homodyning was proposed \cite{Nir2011} and demonstrated \cite{Gross2011}. Homodyning is also feasible via photocurrents induced by signal-pump interference in semiconductors \cite{KurizkiPRB,Kurizki91} and for phonon fields in acoustic structures \cite{khelif_phononic_2015, deymier_acoustic_2013,
kushwaha_acoustic_1993,khelif_transmission_2003,
elnady_quenching_2009}. Any such setup allows to split off a small fraction of the input field and mix it with a correspondingly weak coherent LO, thereby yielding work as per Eqs. (\ref{eq3})-(\ref{effi}). Thus,
the proposed WOF may open new paths towards the exploitation of continuous-variable noise as a source of  useful work in both classical and quantum 
regimes of diverse systems.


{\it Acknowledgement.}-- We thank O. Firstenberg and E. Poem of WIS for useful comments on the manuscript. We acknowledge the support of NSF-BSF, DFG, QUANTERA (PACE-IN), FET-Open (PATHOS) and ISF. T.O. is supported by the Czech Science
Foundation, grant 20-27994S.

\bibliography{wemef}
\begin{widetext}

\appendix*
\setcounter{equation}{0}
\pagenumbering{roman}
\setcounter{page}{1}

\section{Supplemental Material\\ Work Generation from Thermal Noise by Quantum Phase-Sensitive Observation }
\subsection{Work extraction from a quantum harmonic oscillator}
\label{S-work-extract}
\subsubsection{Passivity, ergotropy and work}
{\it A passive state} of a system governed by Hamiltonian $H$ is the state $\Upsilon$ that satisfies
\bn
\mbox{Tr}(\Upsilon H)\leq \mbox{Tr}(U \Upsilon U^\dagger H)
\label{s1}
\en
under any unitary operation $U$ that represents reversible external driving of the system. Inequality (\ref{s1}) implies the 
{\it impossibility of extracting} work from a passive state $\Upsilon$ by unitary operations. For a given $H$, each state $\rho$ has a unique passive counterpart provided both $H$ and $\rho$ are nondegenerate,
\bn
\label{s2}
\Upsilon= U_P \rho U_P^\dagger.
\en
This state is attainable by a unitary transformation $U_P$ which maps the eigenstates of $\rho$ onto the eigenstates of $H$, such that the eigenvalues of $\Upsilon$ monotonically decrease as the corresponding eigenvalues of $H$ increase.

Continuous phase-space distributions, such as the Glauber-Sudarshan P-function used here, are deemed passive if they fall off monotonically and isotropically, from their peak at the origin, as does the Gaussian Gibbs-state distribution (Fig. \ref{Fig1}c).

The mean energy of Hamiltonian $H$ in a state $\rho$ can be decomposed into passive energy $E_{\rm pas}$ and non-passive energy, alias ergotropy, $\mathcal{W}$: 
\bn
\label{s7}
\mbox{tr}(\rho H)=E_{\rm pas}+\mathcal{W}.
\en
The passive energy, which cannot be extracted as useful work in a unitary fashion, is given by
\bn
\label{s4}
E_{\rm pas}=\mbox{Tr}(U_P\rho U_P^\dagger H)=\mbox{Tr}(\Upsilon H),
\en
where $U_P$ is the unitary in Eq. (\ref{s2}) that transforms $\rho$ into its passive counterpart $\Upsilon$.
If $\Upsilon$ is a Gibbs state, then $\Upsilon=Z^{-1}\exp(-H/k_B T)$ and $E_{\rm pas}$ is then thermal energy. 

The term $\mathcal{W}$ in Eq. (\ref{s7}) is
{\it ergotropy}: the maximum amount of work extractable from $\rho$ by a unitary transformation. It is defined as 
\bn
\label{s3}
\mathcal{W}(\rho,H)\equiv \mbox{Tr} (\rho H)- \min_{U}\mbox{Tr}(U \rho U^\dagger H)\geq 0,
\en
the minimization extending over all possible unitary transformations.

The system ergotropy may increase in a non-unitary fashion due to a measurement of an initially passive state,
\bn
\label{s8}
\rho(0)=\Upsilon~\overrightarrow{meas}~\rho(t>0)\neq\Upsilon. 
\en
The corresponding distribution is non-passive if either its peak is {\it displaced from the origin} or its fall-off is {\it anisotropic} and/ or non-monotonic (Fig. \ref{Fig1}c). 
The ergotropy stored in this non-passive state can be subsequently extracted from the system as work via a suitable unitary process (Fig. \ref{Fig1}c),
\bn
\label{s9}
W=\mathcal{W}(\rho(t>0)).
\en
\subsubsection{Mean work and its fluctuations from a displaced harmonic-oscillator state}

Assume a state $\hat \varrho$ such that its mean quadratures are 
\begin{eqnarray}
\label{a1}
\langle \hat x\rangle = {\rm Tr} (\hat \varrho \hat x) = x_0, \\
\langle \hat p\rangle = {\rm Tr} (\hat \varrho \hat p) = p_0 .
\label{a2}
\end{eqnarray}
Let the state be displaced (downshifted to the origin) by a displacement operator $\hat D(-x_0,-p_0) = \exp(-p_0 \hat x+x_0 \hat p) = \hat D^{\dag}(x_0,p_0)$ so that we get the state $\hat \varrho_0$,
\begin{eqnarray}
\hat \varrho_0 = \hat D(-x_0,-p_0)\hat \varrho \hat D^{\dag}(-x_0,-p_0)
\end{eqnarray} 
with zero mean quadratures, i.e.,
\begin{eqnarray}
\label{Etr1}
{\rm Tr} (\hat \varrho_0 x) = 0, \\
{\rm Tr} (\hat \varrho_0 p) = 0. 
\label{Etr2}
\end{eqnarray} 
By this displacement, mean extracted work is equal to the difference of the mean energies of states $\hat \varrho$ and $\hat \varrho_0$,
\begin{eqnarray}
\label{s6}
\langle W\rangle = {\rm Tr} (\hat \varrho \hat H) - {\rm Tr} (\hat \varrho_0 \hat H).
\end{eqnarray}
For a harmonic oscillator
\begin{eqnarray}
\hat H = \frac{1}{2}\left(\hat x^2 + \hat p^2 \right),
\end{eqnarray} 
we obtain from Eq. (\ref{s6})
\begin{eqnarray}
\langle W\rangle &=& \frac{1}{2}{\rm Tr}\left[\hat \varrho (\hat x^2 + \hat p^2) - \hat \varrho_0 (\hat x^2 + \hat p^2)  \right] \nonumber \\
&=& \frac{1}{2}{\rm Tr}\left[\hat \varrho (\hat x^2 + \hat p^2) - \hat D(-x_0,-p_0)\hat \varrho \hat D(x_0,p_0)(\hat x^2 + \hat p^2)  \right] 
\nonumber \\
&=& \frac{1}{2}{\rm Tr}\left[\hat \varrho (\hat x^2 + \hat p^2) - \hat \varrho \hat D(x_0,p_0)(\hat x^2 + \hat p^2)\hat D(-x_0,-p_0)  \right] 
\nonumber \\
&=& \frac{1}{2}{\rm Tr}\left\{\hat \varrho (\hat x^2 + \hat p^2) - \hat \varrho \left[ (\hat x-x_0)^2 + (\hat p-p_0)^2\right]  \right\} 
\nonumber \\
&=& \frac{1}{2}{\rm Tr}\left[\hat \varrho (\hat x^2 + \hat p^2) - \hat \varrho \left(\hat x^2-2x_0 \hat x + x_0^2 + \hat p^2 -2p_0 \hat p + p_0^2\right)  \right] 
\nonumber \\
&=& \frac{1}{2}{\rm Tr}\left[\hat \varrho \left(2x_0 \hat x - x_0^2 + 2p_0 \hat p - p_0^2\right)  \right] 
\nonumber \\
&=& \frac{1}{2} (x_0^2 + p_0^2),
\label{s8}
\end{eqnarray}
where the last line follows from Eqs. (\ref{a1}) and (\ref{a2}).

According to Ref. \cite{Hanggi2007} work is not an observable and its moments should be expressed by means of a characteristic function
\begin{eqnarray}
\mathcal{G}(u) = \int dW \exp (i u W) p(W).
\end{eqnarray}
Here we show that this approach leads to the same result as Eq. (\ref{s8}). Following the approach of Ref. \cite{Hanggi2007}, we start from state $\hat \varrho_0$ and perform work to suddenly change the Hamiltonian from  $\hat H$ to 
\begin{eqnarray}
\hat H'= \frac{1}{2}\left[(\hat x-x_0)^2 + (\hat p - p_0)^2 \right] ,
\end{eqnarray}
without changing the state. 
For the sudden switch from $\hat H$ to $\hat H'$, the characteristic function is 
\begin{eqnarray}
\mathcal{G}(u) = {\rm Tr} \left( e^{iu\hat H'} e^{-iu\hat H}\hat \varrho_0 \right) .
\end{eqnarray}
Expanding the exponentials up to the second order in $u$, one gets
\begin{eqnarray}
\mathcal{G}(u) &\approx& {\rm Tr}\left[\left(1 + iu \hat H' - \frac{u^2}{2}\hat H^{\prime 2} \right)
\left(1 - iu \hat H - \frac{u^2}{2}\hat H^2 \right)\hat \varrho_0 \right] \nonumber \\
&\approx& {\rm Tr}\left\{\left[ 1 + iu (\hat H' - \hat H)
- \frac{u^2}{2}\left(\hat H^{\prime 2} - 2 \hat H' \hat H + \hat H^2  \right) \right]\hat \varrho_0 \right\}.
\end{eqnarray}
The first moment of $W$ is thus
\begin{eqnarray}
\langle W \rangle &=& {\rm Tr}\left[\left( \hat H' - \hat H \right)\hat \varrho_0 \right] \nonumber \\
&=& \frac{1}{2}{\rm Tr}\left[\left( x_0^2+p_0^2 -2x_0\hat x - 2p_0 \hat p \right)\hat \varrho_0 \right] \nonumber \\
&=& \frac{1}{2}{\rm Tr}\left[\left( x_0^2+p_0^2\right)\hat \varrho_0 \right] = \frac{1}{2}(x_0^2 + p_0^2),
\end{eqnarray}
{\it which is the same} as Eq. (\ref{s8}).

The second moment of $W$ is
\begin{eqnarray}
\label{EW}
\langle W^2 \rangle = {\rm Tr}\left[ \left(\hat H^{\prime 2} - 2 \hat H' \hat H + \hat H^2  \right) \hat \varrho_0 \right] .
\end{eqnarray}
Since $\hat H$ and $\hat H'$ do not commute, this expression is in general different from the second moment of the operator $\hat H' - \hat H$,
\begin{eqnarray}
\label{Edif}
\left\langle (\hat H' - \hat H)^2 \right\rangle = {\rm Tr}\left[ \left(\hat H^{\prime 2} - \hat H' \hat H - \hat H \hat H' + \hat H^2  \right) \hat \varrho_0 \right] ,
\end{eqnarray}
This demonstrates the difference between work and the operator  $\hat H' - \hat H$, which was the point of Ref. \cite{Hanggi2007}. The difference between (\ref{EW}) and (\ref{Edif}) is the mean value of the commutator ${\rm Tr} \left( \left[\hat H', \hat H\right] \hat \varrho_0 \right)$.
Since the commutator is
\begin{eqnarray}
\left[\hat H', \hat H\right] = i (p_0 \hat x - x_0 \hat p)
\end{eqnarray}
and because of (\ref{Etr1}) and (\ref{Etr2}), we have
\begin{eqnarray}
{\rm Tr} \left( \left[\hat H', \hat H\right] \hat \varrho_0 \right) = 0,
\end{eqnarray}
\begin{eqnarray}
\label{s19}
\langle W^2 \rangle = \left\langle (\hat H' - \hat H)^2 \right\rangle .
\end{eqnarray}
Thus, we can calculate the first and second moments of extracted work either as the moments of the difference of the original and displaced Hamiltonian (as per Eq. (\ref{s19})) or, equivalently, as following from state displacement per Eq. (\ref{s8}). In what follows, {\it we denote $\langle W \rangle$ as $W$.}



\subsection{Work extraction based on a homodyne measurement}

\subsubsection{Homodyne-measured distribution}
In the setup of Fig. \ref{Fig1}b of the main text, 
photodetection of coherent states $|\gamma_\pm\rangle$,  $|\tilde \gamma_\pm\rangle$ yields Poissonian statistics with mean values 
\begin{eqnarray}
\bar n_{\pm} &=& |\gamma_\pm|^2=\frac{1-\kappa^2}{8}\left[ \left(x \pm\frac{2\beta}{\sqrt{1-\kappa^2}} \right)^2 + p^2 \right] , \\
\bar{\tilde{n}}_{\pm} &=&|\tilde \gamma_\pm|^2=\frac{1-\kappa^2}{8}\left[ x^2 \pm \left( p+\frac{2\beta}{\sqrt{1-\kappa^2}} \right)^2 \right] . \\
\end{eqnarray}
The photocount differences of the two ports of detectors in Fig. \ref{Fig1}b, $\Delta n_x \equiv  n_+-n_-$ and  $\Delta n_p \equiv \tilde {n}_+-\tilde{n}_-$, carry information on the quadratures of the input coherent component $|\alpha\rangle$ with complex amplitude $\alpha =x+ip$. In particular, their mean values are
\begin{eqnarray}
\langle \Delta n_{x} \rangle &=&  \sqrt{1-\kappa^2} \beta x, \\
\langle \Delta n_{p} \rangle &=&  \sqrt{1-\kappa^2} \beta p,
\end{eqnarray}
and their variances are
\begin{eqnarray}
\label{s31}
\langle \Delta n_{x}^2 \rangle - \langle \Delta n_{x} \rangle^2 =
\langle \Delta n_{p}^2 \rangle - \langle \Delta n_{p} \rangle^2 = \frac{1-\kappa^2}{4}
(x^2+p^2) + \beta^2.
\end{eqnarray}
The first and  second terms on the r.h.s. of Eq. (\ref{s31}) are the respective contributions of the split-off field and the LO (Fig. \ref{Fig1}b) to the detected variances which obey the Heisenberg minimal uncertainty.

The probability distribution for the photocount differences can be expressed as the Skellam distribution (distribution of the difference of two statistically independent variables, each with a Poissonian distribution) \cite{Skellam},
\begin{eqnarray}
P(\Delta n_{x}|\alpha) &=& e^{-\bar n_1 - \bar n_2} \left( \frac{\bar n_1}{\bar n_2} \right) ^{\Delta n_{x}/2}
I_{\Delta n_{x}} (2\sqrt{\bar n_1 \bar n_2}) , \\
P(\Delta n_{p}|\alpha) &=& e^{-\bar n_3 - \bar n_4} \left( \frac{\bar n_3}{\bar n_4} \right) ^{\Delta n_{p}/2}
I_{\Delta n_{p}} (2\sqrt{\bar n_3 \bar n_4}) ,
\end{eqnarray}
where $I_k(z)$ is the modified Bessel function of the first kind. The complex amplitude $\alpha$ enters these equations through the dependence of $\bar n_j$ on $x$ and $p$. Upon multiplying these functions, we get the conditional probability distribution for $\Delta n_{x}$ and $\Delta n_{p}$ on condition that the input state was a coherent state $|\alpha\rangle$,
\begin{eqnarray}
 P(\Delta n_{x},\Delta n_{p}|\alpha) = P(\Delta n_{x}|\alpha)P(\Delta n_{p}|\alpha).
\end{eqnarray}

\begin{figure}
\begin{center}
 \includegraphics[width=0.45\linewidth]{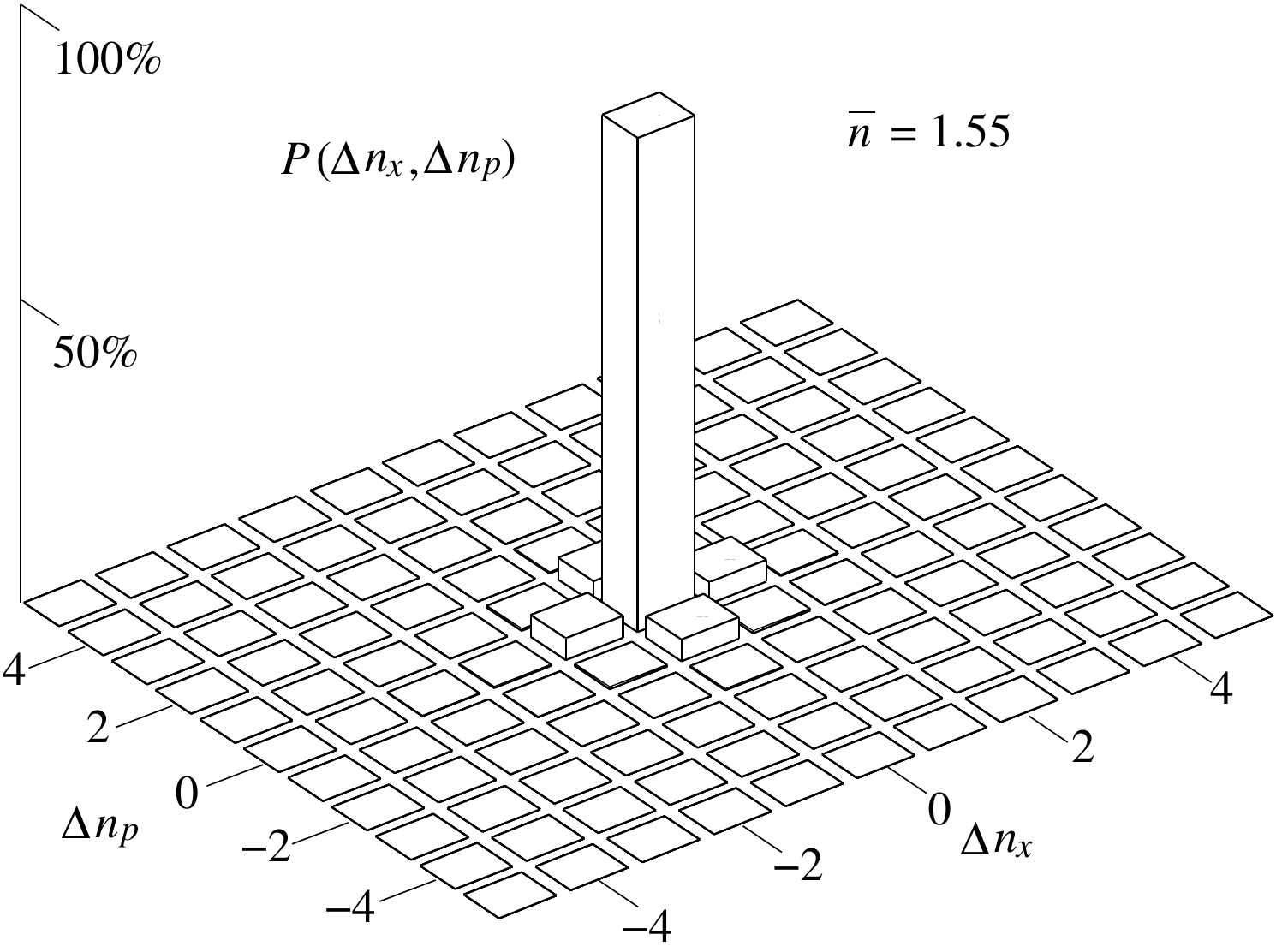}
 \includegraphics[width=0.45\linewidth]{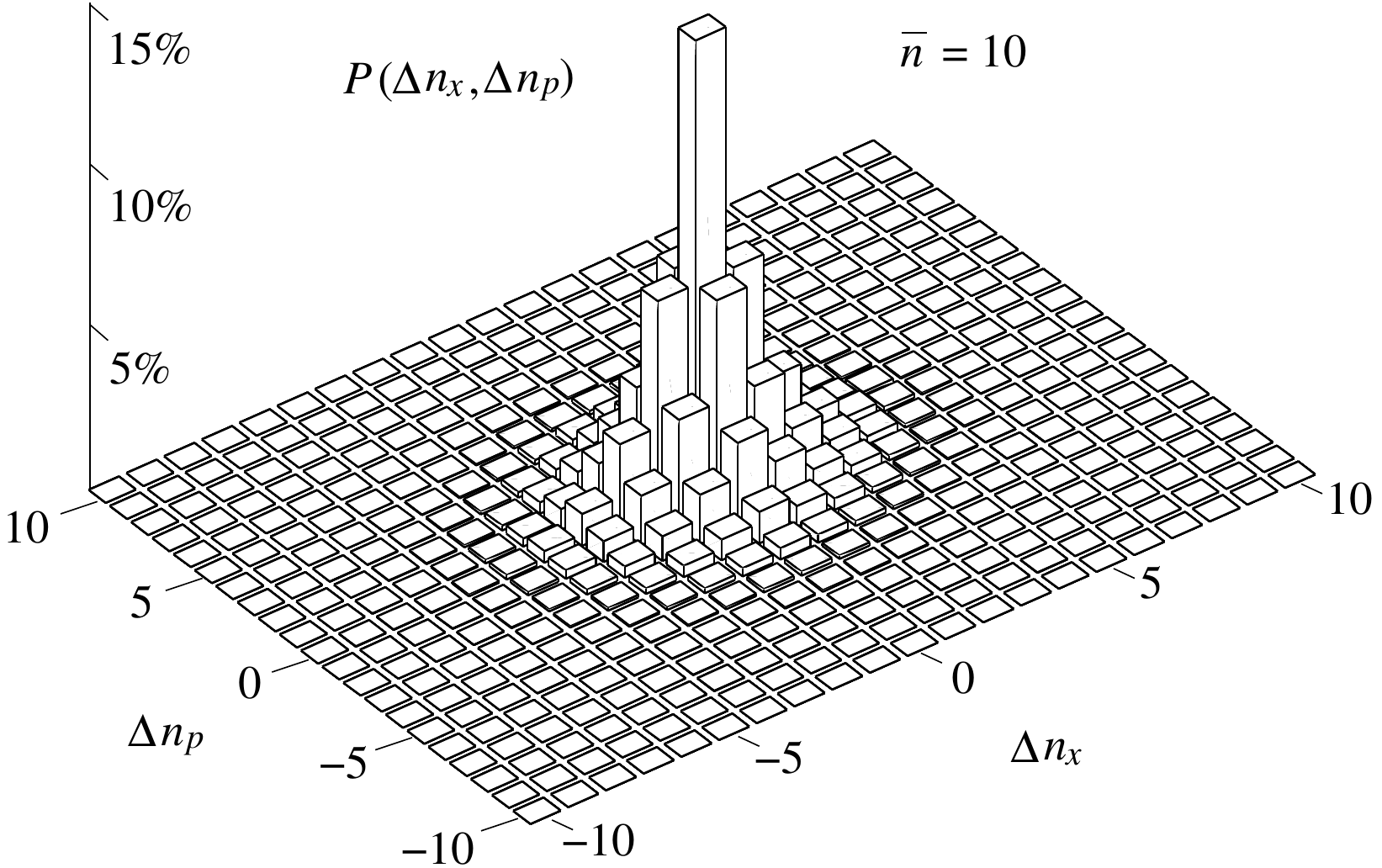}

\caption{\label{f-Fce1}
Probability distribution of detecting photon number differences $\Delta n_x$ and $\Delta n_p$, Eq. (\ref{EqPn}) for $\bar n=1.55$  and $\bar n=10$ with optimized values of $\beta$ and $\kappa$.
}\end{center}
\end{figure}

 Let us now write an arbitrary input state in the basis of coherent states,
\begin{eqnarray}
\hat \varrho = \int \!  \int P(\alpha) |\alpha \rangle \langle \alpha | d^2 \alpha ,
\end{eqnarray}
where $P(\alpha)$ is the Glauber-Sudarshan function corresponding to the input state.
For this state,
the probability distribution of photodetection outcomes is then
\begin{eqnarray}
 P(\Delta n_{x},\Delta n_{p})= \int \!  \int  P(\Delta n_{x},\Delta n_{p}|\alpha) P(\alpha)  d^2 \alpha .
\label{EqPn}
\end{eqnarray}
Examples of the distribution are given in Fig. \ref{f-Fce1}.
For a thermal state with mean number of photons $\bar n$ this function is Gaussian
\begin{eqnarray}
P(\alpha) &=&  \frac{1}{\pi \bar n}\exp \left(- \frac{|\alpha|^2}{\bar n}  \right), \nonumber \\
P(x,p) &=& \frac{1}{2\pi \bar n}\exp \left(- \frac{x^2+p^2}{2\bar n}  \right).
\end{eqnarray}

\begin{figure}

\begin{center}
 \includegraphics[width=0.45\linewidth]{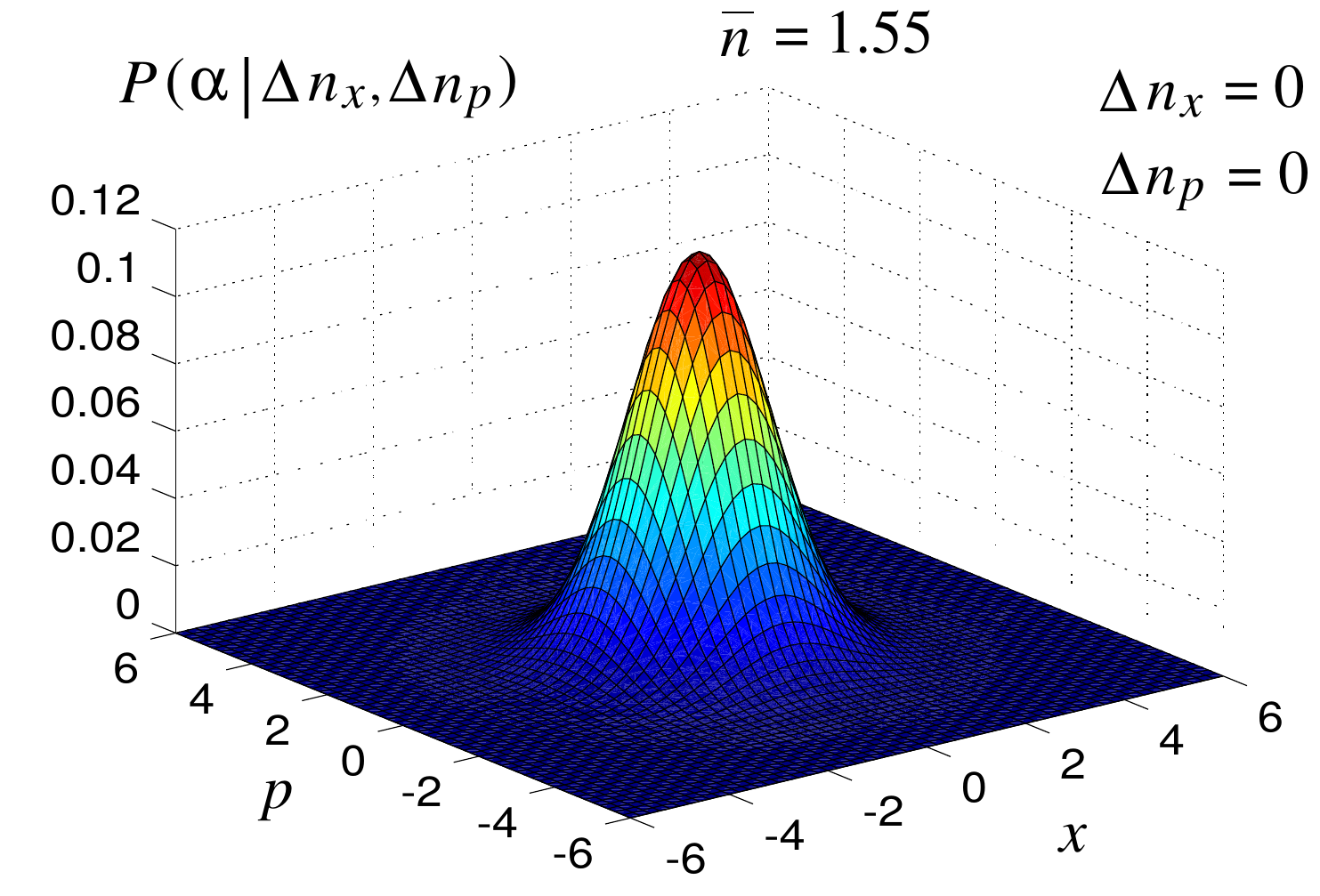}
 \includegraphics[width=0.45\linewidth]{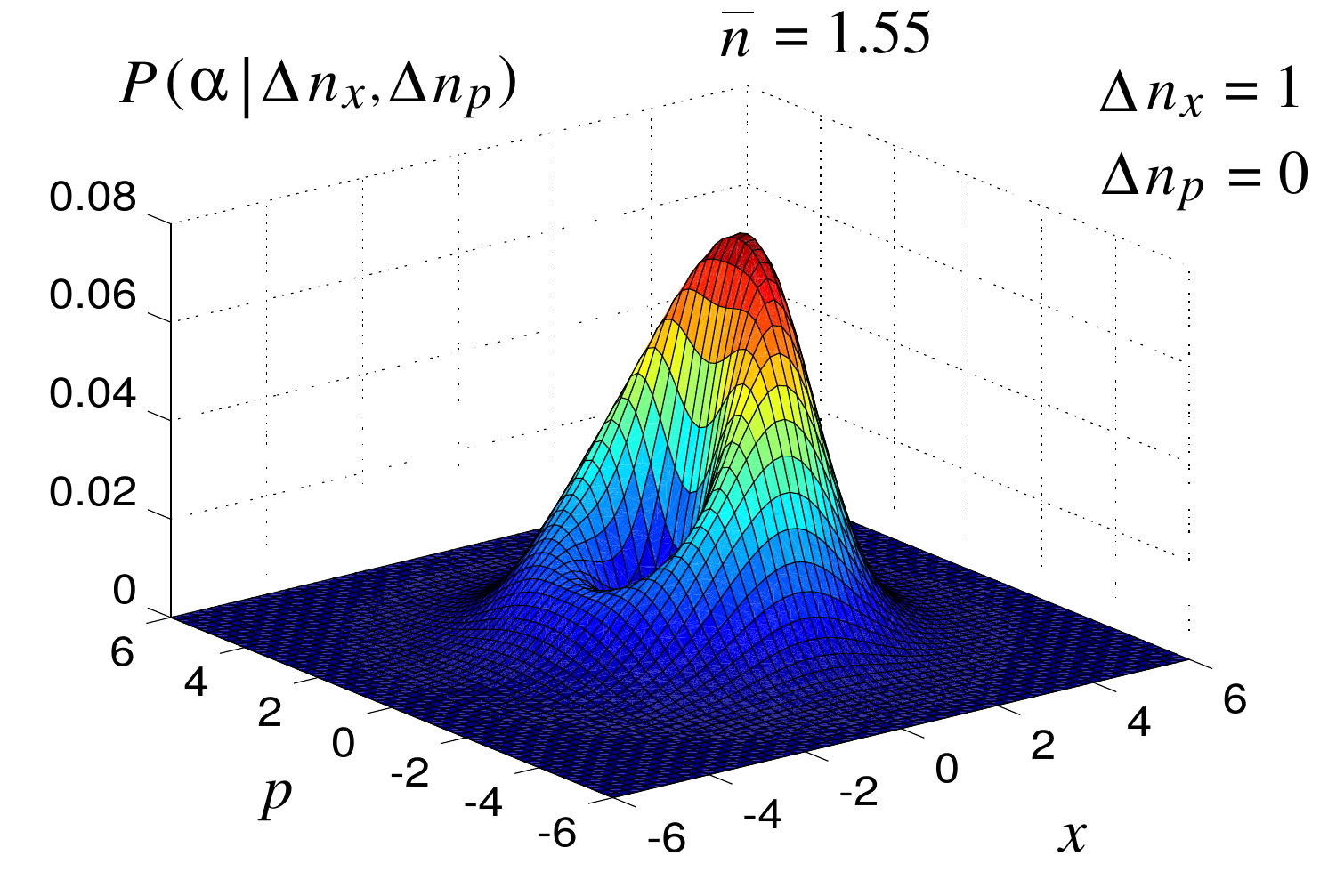}
\caption{\label{f-Fce3}
Examples of the conditional probability of the field quadratures $x,p$ on condition of detected photon number differences $\Delta n_x$ and $\Delta n_p$, Eq. (\ref{PalcondDn}) with $\bar n =1.55$ and optimized values of $\beta =0.113$ and $\kappa=0.9424$.
}\end{center}
\end{figure}

\begin{figure}
\begin{center}
 \includegraphics[width=0.45\linewidth]{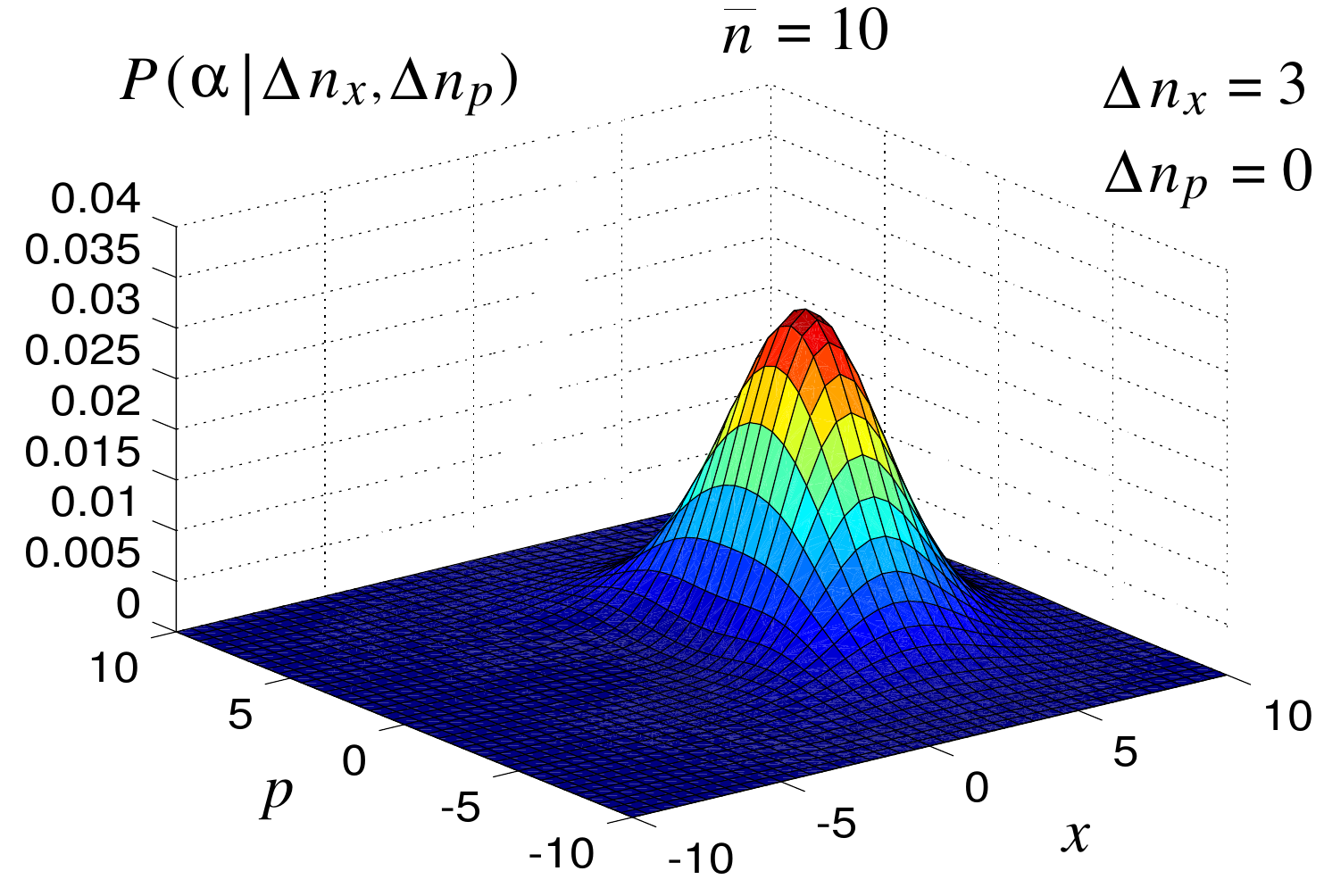}
 \includegraphics[width=0.45\linewidth]{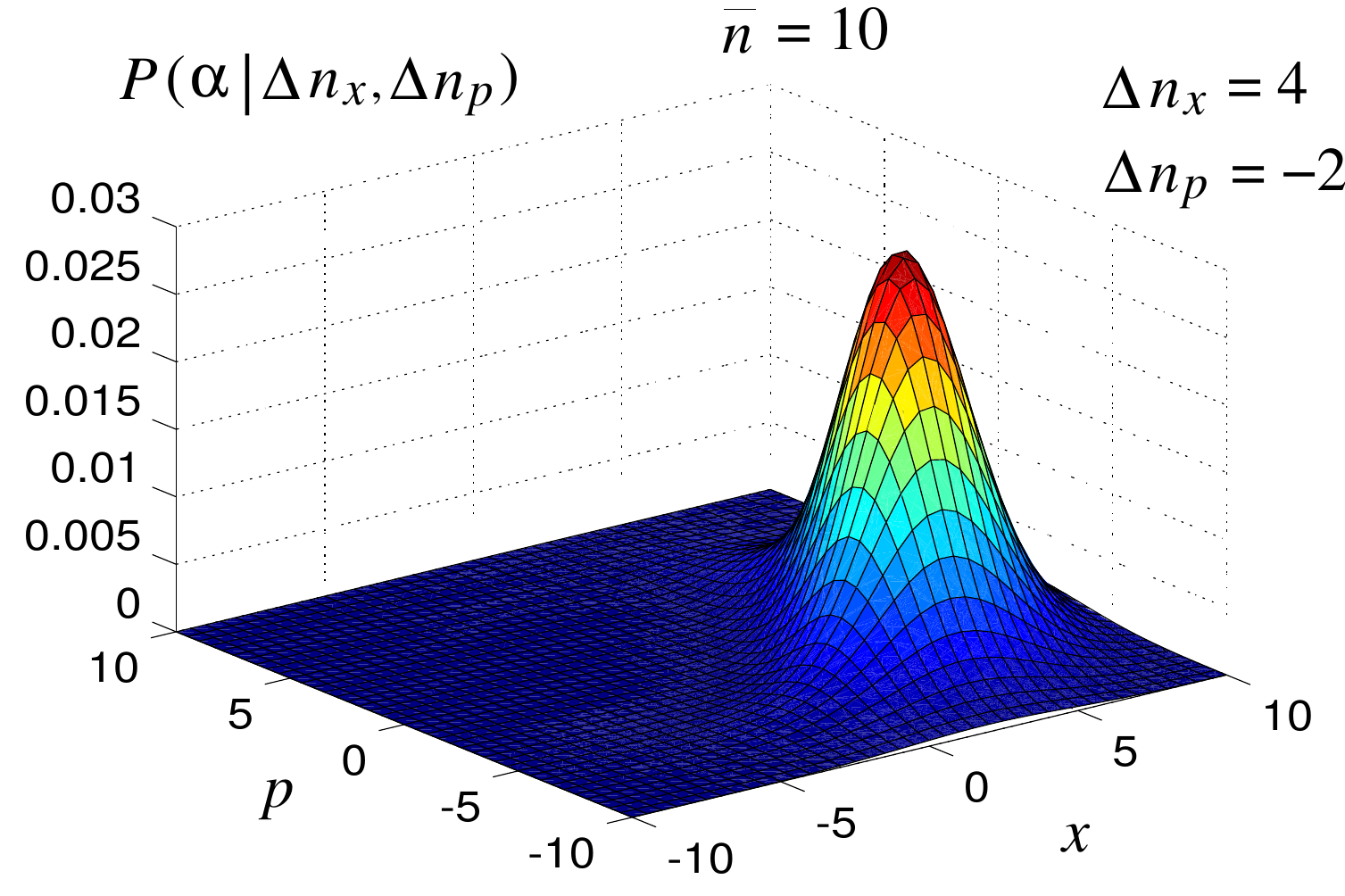}

\caption{\label{f-Fce4}
Conditional probability of the field quadratures $x,p$ on condition of detected photon number differences $\Delta n_x$ and $\Delta n_p$, Eq. (\ref{PalcondDn}) with $\bar n =10$ and optimized values of $\beta =0.780$ and $\kappa=0.902$.
}\end{center}
 
\end{figure}

Upon inverting  relation (\ref{EqPn}) by the Bayes rule, one finds the conditional distribution of $\alpha$ on condition that photon number differences  $\Delta n_{x}$ and $\Delta n_{p}$ were detected as
\begin{eqnarray}
 P(\alpha |\Delta n_{x},\Delta n_{p}) = \frac{P(\Delta n_{x},\Delta n_{p}|\alpha) P(\alpha) }{ P(\Delta n_{x},\Delta n_{p})}.
\label{PalcondDn}
\end{eqnarray}
Examples of these functions for various $\Delta n_{x}$ and $\Delta n_{p}$ are shown in Figs. \ref{f-Fce3} and \ref{f-Fce4}.
The output state (conditional on the detections  $\Delta n_{x}$ and $\Delta n_{p}$) can be expressed as
\begin{eqnarray}
\hat \varrho (\Delta n_{x},\Delta n_{p}) &=& \int \! \int P(\alpha|\Delta n_{x},\Delta n_{p}) |\kappa \alpha \rangle \langle \kappa \alpha | d^2 \alpha  \nonumber \\
&=& \frac{1}{\kappa^2} \int \! \int P\left(\frac{\alpha}{\kappa}|\Delta n_{x},\Delta n_{p}\right) |\alpha \rangle \langle \alpha | d^2 \alpha 
,
\label{varrhoDn}
\end{eqnarray}
which is generally a {\it nonpassive} and non-Gaussian state, 
allthough we have started from a Gaussian  $P(\alpha)$,  as $\alpha$ is not only in the exponential of a  quadratic function.

\subsubsection{Mean extractable work and its variance}
The state corresponding to the detected values $\Delta n_{x}$, $\Delta n_{p}$ has the mean quadratures values
\begin{eqnarray}
\label{eqmeanX}
\langle x\rangle &=& {\rm Tr\ }\left[ \hat x \hat \varrho (\Delta n_{x},\Delta n_{p}) \right]
= \kappa \int \! \int x P(\alpha|\Delta n_{x},\Delta n_{p})  d^2 \alpha , \\
\langle p\rangle &=& {\rm Tr\ }\left[ \hat p \hat \varrho (\Delta n_{x},\Delta n_{p}) \right]
= \kappa \int \! \int p P(\alpha|\Delta n_{x},\Delta n_{p})  d^2 \alpha .
\label{eqmeanP}
\end{eqnarray}
By displacing (downshifting) this state such that the mean quadratures of the resulting state are zero, one gets the work (SM-1) (in units of $\hbar \omega$)
\begin{eqnarray}
W(\Delta n_{x}, \Delta n_{p})=\frac{1}{2}\left(\langle x\rangle ^2 + \langle p\rangle ^2 \right).
\label{WdeltaNxNp}
\end{eqnarray}

The mean net work obtained in this process can be found by averaging this expression over all values of $\Delta n_{x}$, $\Delta n_{p}$ and subtracting the invested energy of the two local oscillators (LO)
which is Eq.  (2) of the main text.
Individual components $(\Delta n_{x}, \Delta n_{p})$ of the extractable work are shown in Fig. \ref{f-Fce2}.



\begin{figure}

 \begin{center}
 \includegraphics[width=0.45\linewidth]{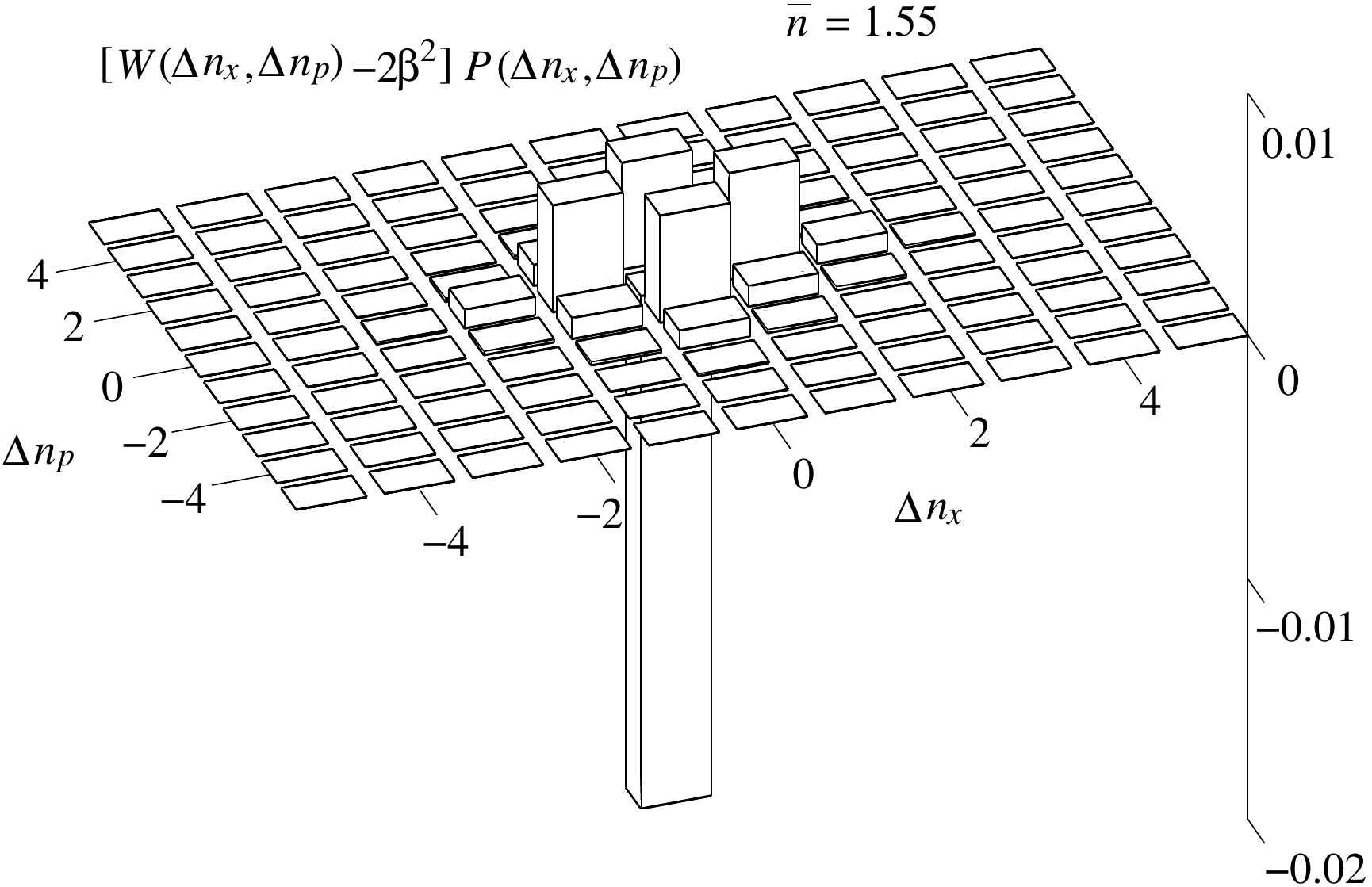}
 \includegraphics[width=0.45\linewidth]{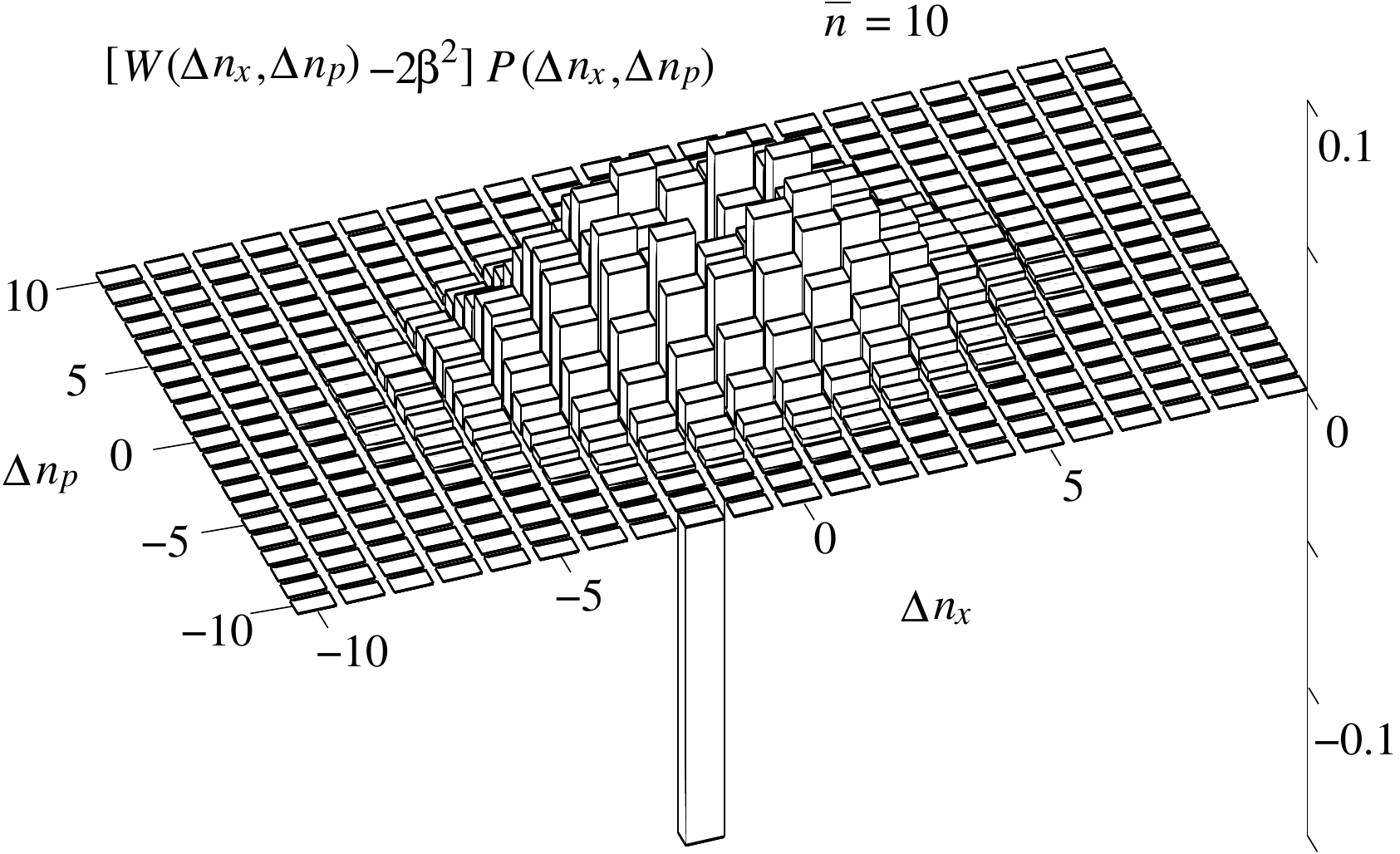}

\caption{\label{f-Fce2}
Product of the probability and work, $[W (\Delta n_{x}, \Delta n_{p}) - 2\beta^2]
P(\Delta n_{x},\Delta n_{p})$ the sum of which is the mean work extractable. In this case the mean extractable work is $W = 0.00447$ for $\bar n=1.55$ and $W = 2.31$ for $\bar n=10$ (in units of $\hbar \omega$).
}\end{center}
\end{figure}
The rms deviation (variance) of the extracted work is given by
\begin{eqnarray}
\Delta W = \sqrt{\overline{W^2}-W^2},
\label{EWfluc}
\end{eqnarray}
where
\begin{eqnarray}
\overline{W^2} = \sum_{\Delta n_{x}} \sum_{\Delta n_{x}}W^2 (\Delta n_{x}, \Delta n_{p})
P(\Delta n_{x},\Delta n_{p})  .
\end{eqnarray}
Note: {\it Throughout the SM we set $\hbar=\omega=1.$}

\subsubsection{Extractable work in the Gaussian approximation}

Assuming that the photon numbers are large enough so that their probability distributions can be approximated by Gaussians, we can write 
\begin{eqnarray}
 P(\Delta n_{x},\Delta n_{p}|\alpha) \approx \frac{1}{2\pi \left[ \frac{(1-\kappa^2)|\alpha|^2}{2}+ \beta^2 \right]}
\exp \left[ {-\frac{\left(\Delta n_{x} - \sqrt{2(1-\kappa^2)}\beta\ {\rm Re\ }\alpha\right)^2
+ \left(\Delta n_{p} - \sqrt{2(1-\kappa^2)}\beta\ {\rm Im\ }\alpha\right)^2}{(1-\kappa^2)|\alpha|^2+2\beta^2}} \right] .
\label{EqPnalpha}
\end{eqnarray}
One can further approximate formula (\ref{EqPnalpha}) by replacing $|\alpha|^2$ with $\bar n$, i.e.,
\begin{eqnarray}
 P(\Delta n_{x},\Delta n_{p}|\alpha) \approx \frac{1}{2\pi \left[ \frac{(1-\kappa^2)\bar n}{2}+ \beta^2 \right]}
\exp \left[ {-\frac{\left(\Delta n_{x} - \sqrt{2(1-\kappa^2)}\beta\ {\rm Re\ }\alpha\right)^2
+ \left(\Delta n_{p} - \sqrt{2(1-\kappa^2)}\beta\ {\rm Im\ }\alpha\right)^2}{(1-\kappa^2)\bar n+2\beta^2}} \right] ,
\label{EqPnalpha2}
\end{eqnarray}
which  allows for analytical integration of Eq. (\ref{EqPn}). One thus gets the Gaussian
\begin{eqnarray}
 P(\Delta n_{x},\Delta n_{p})\approx 
\frac{1}{2\pi \sigma_{\Delta n}^2} \exp \left[ -\frac{\Delta n_{x}^2+\Delta n_{p}^2}{2\sigma_{\Delta n}^2}  \right],
\label{EqPn2}
\end{eqnarray}
whose exact variance is
\begin{eqnarray}
\sigma_{\Delta n}^2 = \beta^2 + \bar n (1-\kappa^2)\left( \beta^2 + \frac{1}{2} \right).
\label{SigmaDeltaN}
\end{eqnarray}



On using Eqs. (\ref{EqPnalpha2}) and (\ref{EqPn2}) in (\ref{PalcondDn}) one finds
\begin{eqnarray}
P(x,p |\Delta n_{x},\Delta n_{p}) 
\approx \frac{1}{2\pi \sigma_x^2} 
\exp \left[-\frac{(x-\bar x_{\Delta nx})^2 + (p-\bar p_{\Delta np})^2}{2\sigma_x^2} \right],
\label{S49}
\end{eqnarray}
where
\begin{eqnarray}
\label{ApproxBarx}
\bar x_{\Delta nx} &=& \frac{ \Delta n_x}{\beta \sqrt{1-\kappa^2}\left[1+\frac{1}{\bar n (1-\kappa^2)}+\frac{1}{2\beta^2} \right]} , \\
\label{ApproxBarp}
\bar p_{\Delta np} &=& \frac{ \Delta n_p}{\beta \sqrt{1-\kappa^2}\left[1+\frac{1}{\bar n (1-\kappa^2)}+\frac{1}{2\beta^2} \right]} , \\
 \sigma_x^2 &=& \frac{\bar n}{1+\frac{2\beta^2\bar n (1-\kappa^2)}{2\beta^2+\bar n (1-\kappa^2)}} .
\label{SigmaX}
\end{eqnarray}
Using (\ref{ApproxBarx}) and (\ref{ApproxBarp}) to approximate $\langle x\rangle$ and  $\langle p\rangle$ in (\ref{WdeltaNxNp}) as
\begin{eqnarray}
\langle x \rangle &\approx& \kappa \bar x_{\Delta nx} , \\
\langle p \rangle &\approx& \kappa \bar p_{\Delta np} ,
\end{eqnarray}
and $P(\Delta n_{x},\Delta n_{p})$ from Eq. (\ref{EqPn2}) one finds for the mean work
\begin{eqnarray}
W_{\rm} &\approx & \frac{1}{2}\int \int 
\left(\langle x \rangle^2 + \langle p \rangle^2  \right)
P(\Delta n_{x},\Delta n_{p}) d\Delta n_{x} \ d\Delta n_{p} -2\beta^2 \nonumber \\
&\approx& \frac{2\beta^2\kappa^2 (1-\kappa^2) \bar n^2}{2\beta^2 +(1-\kappa^2)(1+2\beta^2)\bar n} -2\beta^2 .
\label{WnetApprox}
\end{eqnarray}
Equation (\ref{WnetApprox}) turns out, according to numerical checks, to be a very good approximation for all values of $\bar n$. 
After extracting the work, the remaining state is thermal, with energy 
\begin{eqnarray}
E_{\rm rem} = \kappa^2  \sigma_x^2 .
\label{Erest0}
\end{eqnarray}
One can estimate the fluctuations (variance) $\Delta W$ according to (\ref{EWfluc}).
By integrating the Gaussian (\ref{S49}) multiplied by $(\Delta n_x^2+\Delta n_p^2)^2$ one finds
\begin{eqnarray}
\Delta W &\approx &  \frac{2\beta^2\kappa^2 (1-\kappa^2) \bar n^2}{2\beta^2 +(1-\kappa^2)(1+2\beta^2)\bar n}=W+2\beta^2.
\end{eqnarray}



\subsubsection{Extractable work in the low-excitation approximation }
For weakly excited input we can assume that the states arriving at the photodetectors are coherent states with $|\beta|\ll 1$ and $|\sqrt{1-\kappa^2}\alpha| \ll 1$. The Poissonian statistics of photocounts can then be replaced with nonzero values valid only for photon numbers of 0 or 1, the rest being neglected. One finds
\begin{eqnarray}
P(\Delta n_x = 0|x,y) = P(\Delta n_p = 0|x,y) &\approx & 1 -\epsilon \frac{x^2+p^2}{2} - 2\beta^2, \\
P(\Delta n_x = \pm 1|x,y)  &\approx & \epsilon \frac{x^2+p^2}{8} + \frac{\beta^2}{2} \pm \frac{\sqrt{\epsilon}\beta}{2}x, \\
P(\Delta n_p = \pm 1|x,y)  &\approx & \epsilon \frac{x^2+p^2}{8} + \frac{\beta^2}{2} \pm \frac{\sqrt{\epsilon}\beta}{2}p, 
\end{eqnarray}
where $\epsilon =1-\kappa^2$.
Eq. (\ref{EqPn}) leads to
\begin{eqnarray}
P(\Delta n_x=0, \Delta n_p=0) &\approx & 1-2\beta^2 -\epsilon \bar n, \\
P(\Delta n_x=\pm 1, \Delta n_p=0) = P(\Delta n_x=0, \Delta n_p=\pm 1) &\approx & \frac{\beta^2}{2} + \frac{\epsilon \bar n}{4},
\end{eqnarray}
upon neglecting the probabilities of other photocounts.
Inverting the conditional probabilities as in (\ref{PalcondDn}) one finds
\begin{eqnarray}
P(x,p|\Delta n_x=0, \Delta n_p=0) &\approx & \frac{1-2\beta^2 -\frac{\epsilon}{2}(x^2+p^2)}{2\pi \bar n(1-2\beta^2 -\epsilon \bar n)}\exp \left(-\frac{x^2+p^2}{2\bar n} \right), \\
P(x,p|\Delta n_x=\pm 1, \Delta n_p=0) &\approx & \frac{2\beta^2 +\frac{\epsilon}{2}(x^2+p^2)\pm 2\beta \sqrt{\epsilon} x}{2\pi \bar n(2\beta^2 +\epsilon \bar n)}\exp \left(-\frac{x^2+p^2}{2\bar n} \right), \\
P(x,p|\Delta n_x=0, \Delta n_p=\pm 1) &\approx & \frac{2\beta^2 +\frac{\epsilon}{2}(x^2+p^2)\pm 2\beta \sqrt{\epsilon} p}{2\pi \bar n(2\beta^2 +\epsilon \bar n)}\exp \left(-\frac{x^2+p^2}{2\bar n} \right).
\end{eqnarray}
The conditional mean values of the quadratures are
\begin{eqnarray}
\langle x \rangle &=& 0 \qquad {\rm for} \ \Delta n_x =0, \\
\langle x \rangle &=& \pm \frac{2\beta \kappa \sqrt{\epsilon} \bar n}{\epsilon \bar n + 2\beta^2} \qquad {\rm for} \ \Delta n_x =\pm 1, \\
\langle p \rangle &=& 0 \qquad {\rm for} \ \Delta n_p =0, \\
\langle p \rangle &=& \pm \frac{2\beta \kappa \sqrt{\epsilon} \bar n}{\epsilon \bar n + 2\beta^2} \qquad {\rm for} \ \Delta n_p =\pm 1.
\end{eqnarray}
The mean extractable work is then 
\begin{eqnarray}
W &\approx & \frac{2\beta^2\kappa^2 (1-\kappa^2) \bar n^2}{2\beta^2 +(1-\kappa^2)\bar n} -2\beta^2 ,
\label{WnetApprox2}
\end{eqnarray}
which yields results very close to those following from the Gaussian approximation (\ref{WnetApprox}).

\subsection{ Downshift (displacement) Work optimization}
\label{S-work-optimization}


In the Gaussian approximation, one can maximize Eq. (\ref{WnetApprox}) with respect to $\beta$ and $\kappa$ as
follows.
On using the substitution $\xi\equiv 2\beta^2$, and $\epsilon \equiv 1-\kappa^2$, one can write Eq. (\ref{WnetApprox}) as
\begin{eqnarray}
W = \bar n \frac{1-\epsilon}{1+\frac{1}{\xi}+\frac{1}{\epsilon \bar n}}-\xi.
\label{WnAp}
\end{eqnarray}
Setting the derivative with respect to $\xi$ equal to zero, $\partial W/\partial \xi = 0$ leads to a quadratic equation
\begin{eqnarray}
\left(1+ \frac{1}{\epsilon \bar n}\right)^2 \xi^2 + 2 \left(1+ \frac{1}{\epsilon \bar n}\right) \xi
+ 1-\bar n (1-\epsilon) = 0,
\end{eqnarray}
which has one positive root, namely
\begin{eqnarray}
\xi=\frac{\sqrt{\bar n (1-\epsilon)}-1}{1+\frac{1}{\epsilon \bar n}}.
\label{Apxi}
\end{eqnarray}
Inserting this to (\ref{WnAp}), one gets
\begin{eqnarray}
W= \bar n \epsilon \frac{1+\bar n - \epsilon \bar n -2\sqrt{\bar n}\sqrt{1-\epsilon}}{\bar n \epsilon +1}.
\end{eqnarray}
For $\epsilon \ll 1$ (i.e., high transmissivity of the $0$th beam splitter BS0 in Fig. 1b) one can use the linear approximation of $\sqrt{1-\epsilon}\approx 1-\epsilon/2$ to get
\begin{eqnarray}
W \approx \bar n \epsilon \frac{(\sqrt{\bar n}-\bar n)\epsilon + (\sqrt{\bar n}-1)^2}{\bar n \epsilon +1}.
\label{WnAp2}
\end{eqnarray}
Setting the derivative equal to zero, $\partial W/\partial \epsilon =0$ one gets a quadratic equation 
\begin{eqnarray}
\bar n \epsilon^2 + 2 \epsilon - 1+ \frac{1}{\sqrt{\bar n}} =0,
\end{eqnarray}
which has one positive root, namely
\begin{eqnarray}
\epsilon = \frac{\sqrt{\bar n -\sqrt{\bar n}+1}-1}{\bar n}.
\label{Apepsilon}
\end{eqnarray}
Inserting Eq. (\ref{Apepsilon}) into (\ref{WnAp2}) we get Eq. (3) of the main text for maximized work,
\begin{eqnarray}
W_{\rm max} \approx
\left(\sqrt{\bar n - \sqrt{\bar n}+1}-1 \right)^2  \left( 1-\frac{1}{\sqrt{\bar n}} \right),
\label{Woptimum}
\end{eqnarray}
achieved for 
\begin{eqnarray}
\beta = \sqrt{\frac{\xi}{2}}, \qquad
\kappa &=& \sqrt{1-\epsilon},
\label{EOpteta}
\end{eqnarray}
where $\epsilon$ is given by (\ref{Apepsilon}) and $\xi$  by (\ref{Apxi}).
A very similar result is obtained also by optimizing the formula for the low excitation limit.

\begin{figure}

\centering
\includegraphics[width=8cm]{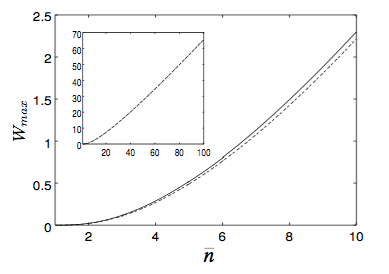}
\caption{
Dependence of $W_{max}$ on $\bar{n}$. Full line: numerical optimization of the exact function in Eq. (2). Dashed line: approximate formula in Eq. (3).
}
\label{supplefig}
\end{figure}

\subsubsection{High-temperature limit}
For high temperatures ($\bar n \gg 1$) one can expand Eq. (3) in the main text as 
\begin{eqnarray}
W_{\rm max} \approx \bar n - 4\sqrt{\bar n} + 6  
\label{EWoptHot}
\end{eqnarray}
In this case the portion of the input energy split off at the first beam splitter is (taking into account only the two largest terms)
\begin{eqnarray}
(1-\kappa^2)\bar n  \approx \sqrt{\bar n} -\frac{3}{2}
\end{eqnarray}
and the LO energy is
\begin{eqnarray}
E_{\rm LO} = 2\beta^2\approx\sqrt{\bar n}-\frac{5}{2} .
\end{eqnarray}
%
The detectors absorb the energy 
\begin{eqnarray}
E_{\rm det} = (1-\kappa^2) \bar n  + E_{\rm LO} \approx 2 \sqrt{\bar n}-4, 
\end{eqnarray}
work $W$ is extracted.
The remaining (unused) output energy
\begin{eqnarray}
E_{\rm rem} \approx 2\sqrt{\bar n}-2,
\label{Erest}
\end{eqnarray}
is thermal in the Gaussian approximation, being associated with the output field fluctuations. Then $W$ in (\ref{EWoptHot}) is found from the energy balance
\begin{eqnarray}
E_{\rm in} + E_{\rm LO} = E_{\rm det} + W +E_{\rm rem} \approx \bar n + \sqrt{\bar n}-\frac{5}{2}.
\end{eqnarray}
It can be deduced from \ref{EOpteta}-\ref{Erest} that consecutive iterations that exploits $E_{\rm rem}$ as the input do not contribute significantly to the work output.

\subsubsection{Low-temperature limit}
For $\bar n$ just barely above 1 one can expand Eq. (3) in terms of $(\bar n-1)$ as 
\begin{eqnarray}
W_{\rm max} \approx \frac{(\bar n -1)^3}{32}.
\label{Wopt32}
\end{eqnarray}
However,  this result follows from the Gaussian approximation which holds for high temperature.
Optimizing Eq. (2) with respect to $\kappa$ and $\beta$, one finds 
\begin{eqnarray}
W_{\rm max} \approx
\frac{\bar n - \sqrt{\bar n}}{2}\left(\sqrt{\frac{\bar n + \sqrt{\bar n}}{2}}-1 \right)^2 ,
\label{Woptimum2}
\end{eqnarray}
achieved for 
\begin{eqnarray}
2\beta^2 &=& \frac{\bar n - \sqrt{\bar n}}{2}\left(\sqrt{\frac{\bar n + \sqrt{\bar n}}{2}}-1\right) , \\
1-\kappa^2 &=& \frac{1-\frac{1}{\sqrt{\bar n}}}{2}.
\end{eqnarray}
Expanding Eq. (\ref{Woptimum2}) for $\bar n$ near 1 leads to a result similar to Eq.(\ref{Wopt32})
\begin{eqnarray}
W_{\rm max} \approx \frac{9}{256}(\bar n -1)^3.
\end{eqnarray}

\subsection{Work extraction by unsqueezing}
\label{S-squeezing}

Even after displacing the output state (\ref{varrhoDn}) to the origin, in general one does not end up with a passive state. Further work extraction can then be achieved, e.g., by means of unsqueezing operations. These operations can be performed by letting the downshifted output interact with a Kerr medium, as in Refs.  \cite{Carmichael_BOOK,Walls_BOOK,Scully_BOOK}.

For the purpose of work extraction, assume first a downshifted state with first moments equal to zero $\langle x \rangle = \langle p \rangle = 0$ and with a variance matrix $V$ with 
\begin{eqnarray}
V_{11} = \langle \hat x^2 \rangle, V_{12} = V_{21} = \frac{1}{2}\langle \hat x\hat p+\hat p\hat x \rangle,  V_{22} = \langle \hat p^2 \rangle
\end{eqnarray}
 The eigenvalues of the variance matrix are the principal variances
 \begin{eqnarray}
 V_{\pm}=\frac{V_{11}+V_{22}}{2}\pm \frac{1}{2}\sqrt{(V_{11}-V_{22})^2+4V_{12}^2} .
 \end{eqnarray}
The energy of the state is $E_0=\frac{1}{2}( \langle \hat x^2 \rangle +\langle \hat p^2 \rangle ) = \frac{1}{2}(V_++V_-)$. Any unsqueezing operation will leave the product of the principal variances $V_+V_-$ unchanged.

Among all states that can be generated by unsqueezing operations, the one with equal principal variances 
\begin{eqnarray}
V_{f+}=V_{f-}=\sqrt{V_+V_-}
\end{eqnarray}
 has the smallest energy 
 \begin{eqnarray}
E_f=\sqrt{V_+V_-}.
\end{eqnarray}
\begin{figure}
\centering
  \includegraphics[width=8cm]{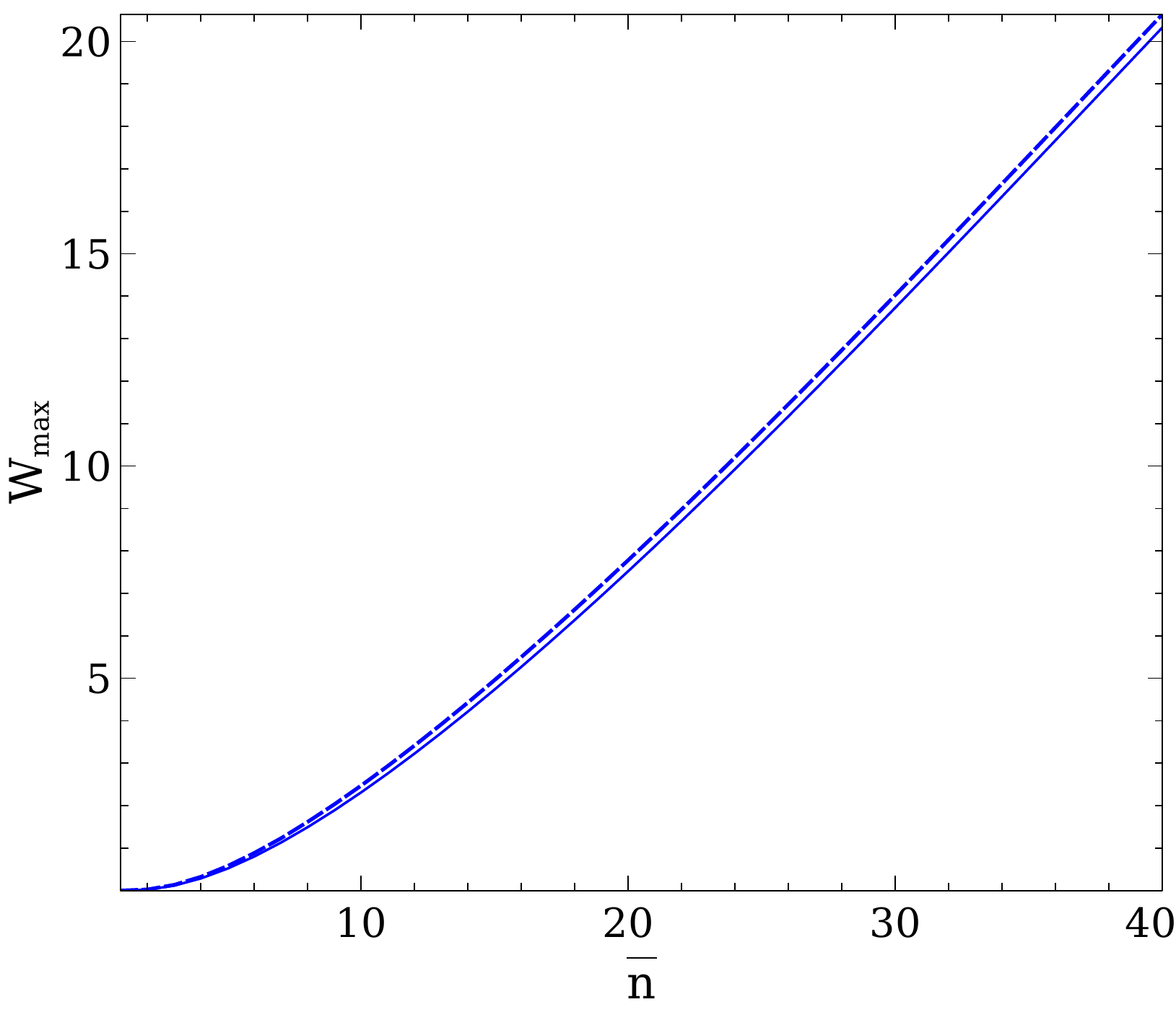}
\caption{Work extraction via displacement and unsqueezing (dashed) compared to its counterpart without unsqueezing (solid) in units of $\hbar \omega$ as a function of the mean input number of quanta $\bar n$.
}
\label{unsqueezing}
\end{figure}
 The difference between the initial and final energies can be extracted as work by the unsqueezing operation, $W_{US} (\Delta n_{x},\Delta n_{p})=E_0-E_f$, where
\begin{eqnarray}
W_{US} (\Delta n_{x},\Delta n_{p}) = \frac{V_++V_-}{2}-\sqrt{V_+V_-} = \frac{V_{11}+V_{22}}{2}-\sqrt{V_{11}V_{22}-V_{12}^2}.
\end{eqnarray}
The elements of the variance matrix can be found from the properties of the Glauber-Sudarshan function as
\begin{eqnarray}
V_{11}   &=& {\rm Tr\ }\left[ \hat x^2 \hat \varrho (\Delta n_{x},\Delta n_{p}) \right] - \langle x\rangle^2
= \kappa^2 \int \! \int x^2 P(\alpha|\Delta n_{x},\Delta n_{p})  d^2 \alpha - \langle x\rangle^2 + \frac{1}{2}, \\
V_{22}   &=& {\rm Tr\ }\left[ \hat p^2 \hat \varrho (\Delta n_{x},\Delta n_{p}) \right] - \langle p\rangle^2
= \kappa^2 \int \! \int p^2 P(\alpha|\Delta n_{x},\Delta n_{p})  d^2 \alpha - \langle p\rangle^2 + \frac{1}{2}, \\
V_{12}  &=& \frac{1}{2}{\rm Tr\ }\left[ (\hat x \hat p + \hat p\hat x)\hat \varrho (\Delta n_{x},\Delta n_{p}) \right] - \langle x\rangle \langle p\rangle
= \kappa^2 \int \! \int xp P(\alpha|\Delta n_{x},\Delta n_{p})  d^2 \alpha - \langle x\rangle \langle p\rangle ,
\end{eqnarray}
where $\langle x\rangle$ and  $\langle p\rangle$ are given by Eqs. (\ref{eqmeanX}) and (\ref{eqmeanP}).
The average work extracted by unsqueezing is then
\begin{eqnarray}
W_{US} = \sum_{\Delta n_{x}} \sum_{\Delta n_{x}}W_{US} (\Delta n_{x}, \Delta n_{p})
P(\Delta n_{x},\Delta n_{p}) .
\label{EWnetS}
\end{eqnarray} 
Numerical results (see main text) show that the contribution
to the extractable work from $W_{US}$ is small; diminishes rapidly with $\bar{n}$, and is negligible compared to work obtained by displacement for 
$\bar n\gg1$ (Fig. \ref{unsqueezing}).

\subsection{ Imperfect photodetection and Spurious Thermal noise}
\label{S-thermal-noise}
In addition to the resetting cost, one must reckon with thermal  noise  from  ``parasitic'' (spurious)  sources  that may be incident on the unused input ports, as well as accompany the LO or give rise to detector dark counts. 
To study and influence of spurious thermal noise at the unused ports, let us consider the scheme as in Fig. \ref{f-heterodyne2c}. The spurious noise coming through the first beam splitter is modeled as an ensemble of coherent states $|\tau \rangle$ with thermal distribution ${\rm P}(\tau)$ with mean photon number $\bar n_{\tau}$. Spurious noise coming through the homodyne beam splitter has mean photon number $n_{\rm H}$. The local oscillators are considered as displaced thermal states, their coherent amplitudes being $\beta$ and $i\beta$, and the mean number of thermal photons in each of them is $\bar n_{\rm LO}$. 

The imperfect photodetectors are modeled as ideal photodetectors with beam splitters in front of them. Each of these beam splitters transmits the fraction $\kappa^2_D$ of the input energy and reflects (diverts) $1-\kappa^2_D$. Thermal light with mean number of photons $\bar n_D$ enters the unused input of each of these beam splitters.

\begin{figure}
\centering
  \includegraphics[width=14cm]{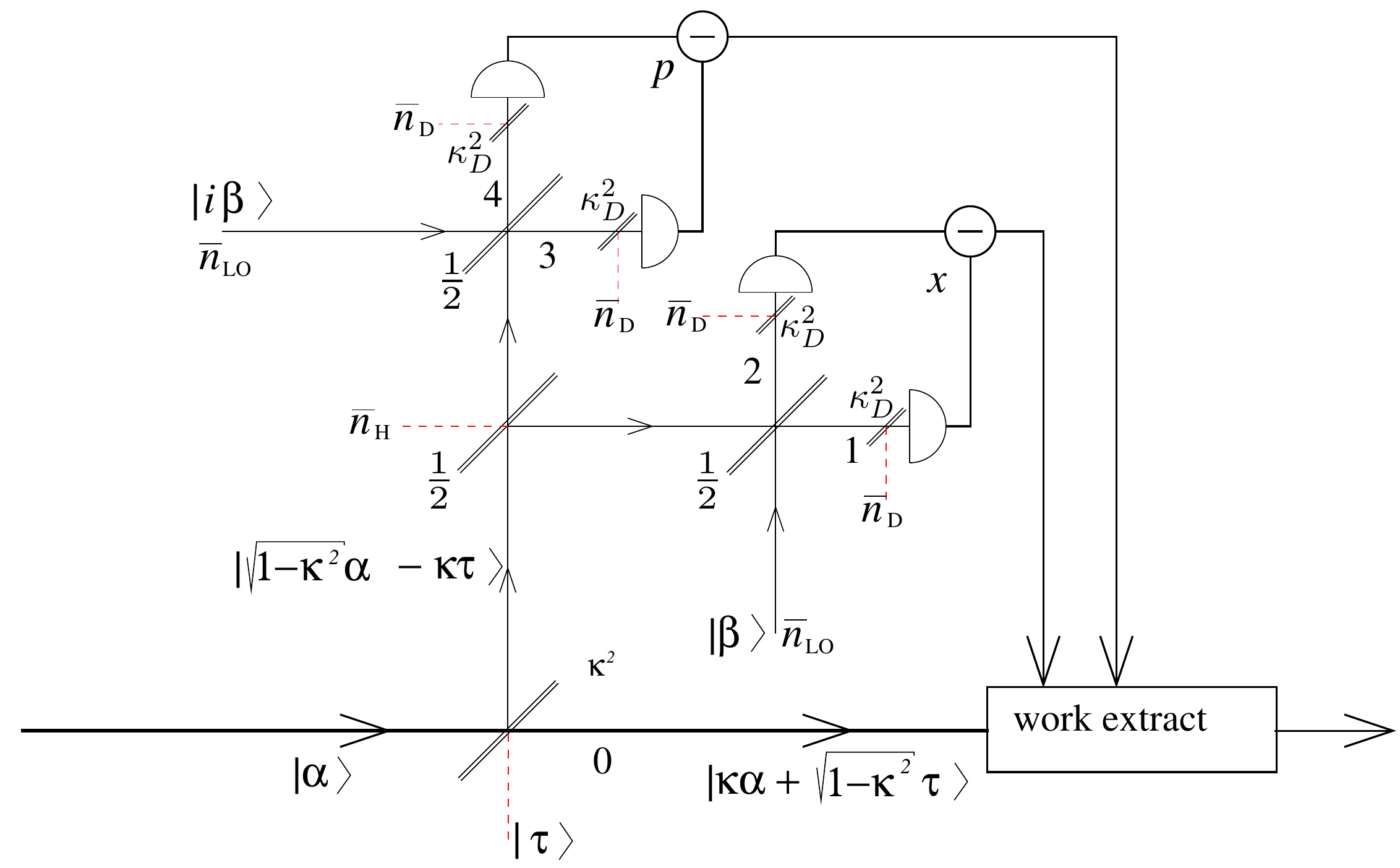}
\caption{
Scheme with spurious (thermal) noise input and imperfect photodetection. Spurious (thermal) noise enters unused ports of the beam splitters, and the local oscillators are modeled by displaced thermal states with $\bar n_{\rm LO}$ photons.}
\label{f-heterodyne2c}
\end{figure}


We need to calculate the conditional values of mean output quadratures provided that photon number differences $\Delta n_x$ and $\Delta n_p$ were detected. We will estimate the values by the Gaussian approximation of the photon number distributions. Assume first that the input states at the first beam splitter are coherent $|\alpha\rangle$ and $|\tau\rangle$ with $\alpha=x+ip$  and
\begin{eqnarray}
\tau = \frac{1}{\sqrt{2}}(x_{\tau}+ ip_{\tau}).
\end{eqnarray}
The state approaching the homodyne detectors is coherent, with quadratures
\begin{eqnarray}
\tilde x &=& \sqrt{1-\kappa^2} x - \kappa x_{\tau}, 
\\
\tilde p &=& \sqrt{1-\kappa^2} p - \kappa p_{\tau}. 
\end{eqnarray}
Using the transformation properties of the beam splitters, one can find the mean numbers of photons arriving at the photodetectors as
\begin{eqnarray}
\bar n_1 &=& \frac{\kappa_D^2}{8}(\tilde x^2 + \tilde p^2)+\frac{\kappa_D^2}{2}\beta^2+\frac{\kappa_D^2}{2}\beta \tilde x
+ \frac{\kappa_D^2}{4}\bar n_{\rm H}+\frac{\kappa_D^2}{2}\bar n_{\rm LO} + (1-\kappa_D^2)\bar n_{\rm D}, \\
\bar n_2 &=& \frac{\kappa_D^2}{8}(\tilde x^2 + \tilde p^2)+\frac{\kappa_D^2}{2}\beta^2-\frac{\kappa_D^2}{2}\beta \tilde x
+ \frac{\kappa_D^2}{4}\bar n_{\rm H}+\frac{\kappa_D^2}{2}\bar n_{\rm LO} + (1-\kappa_D^2)\bar n_{\rm D}, \\
\bar n_3 &=& \frac{\kappa_D^2}{8}(\tilde x^2 + \tilde p^2)+\frac{\kappa_D^2}{2}\beta^2+\frac{\kappa_D^2}{2}\beta \tilde p
+ \frac{\kappa_D^2}{4}\bar n_{\rm H}+\frac{\kappa_D^2}{2}\bar n_{\rm LO} + (1-\kappa_D^2)\bar n_{\rm D}, \\
\bar n_4 &=& \frac{\kappa_D^2}{8}(\tilde x^2 + \tilde p^2)+\frac{\kappa_D^2}{2}\beta^2-\frac{\kappa_D^2}{2}\beta \tilde p
+ \frac{\kappa_D^2}{4}\bar n_{\rm H}+\frac{\kappa_D^2}{2}\bar n_{\rm LO} + (1-\kappa_D^2)\bar n_{\rm D},
\end{eqnarray}
so that for the photon number differences we get
\begin{eqnarray}
 \langle \Delta n_x \rangle  = \langle n_1-n_2 \rangle &=& \kappa_D^2 \beta \tilde x, \\
 \langle \Delta n_p \rangle  =  \langle n_3-n_4 \rangle &=& \kappa_D^2 \beta \tilde p. 
\end{eqnarray}
The second moments of the detected photon number differences are
\begin{eqnarray}
\nonumber
D^2 &\equiv & 
\langle (n_1-n_2)^2 \rangle -\langle n_1-n_2 \rangle^2 = \langle (n_3-n_4)^2 \rangle -\langle n_3-n_4 \rangle^2 \\
\nonumber
&=& 
\kappa_D^4 \bar n_{\rm H}\bar n_{\rm LO}
+ \frac{\kappa_D^2(1-\kappa_D^2)}{2}\left( \bar n_{\rm H} + 2\bar n_{\rm LO} \right) \bar n_{\rm D}
+ 2 (1-\kappa_D^2)^2\bar n_{\rm D}^2  
+ \kappa_D^2\left[1+2 (1-\kappa_D^2)\bar n_{\rm D}\right] \beta^2
 \nonumber \\
& & 
+ \frac{\kappa_D^2}{4}\left[\kappa_D^2 ( \bar n_{\rm H} +2 \bar n_{\rm LO}) + 2(1-\kappa_D^2)\bar n_{\rm D}+1 \right] 
(\tilde x^2 + \tilde p^2)
+ \frac{\kappa_D^2}{2}\bar n_{\rm H} + \kappa_D^2\bar n_{\rm LO} + 2 (1-\kappa_D^2)\bar n_{\rm D} .
\label{EqD2}
\end{eqnarray}
The conditional probability distribution $P(\Delta n_x, \Delta n_p | \alpha, \tau)$ can be inverted as
\begin{eqnarray}
P(\alpha, \tau | \Delta n_x, \Delta n_p) &=& \frac{P(\Delta n_x, \Delta n_p | \alpha, \tau)P(\alpha) P(\tau)}
{P(\Delta n_x, \Delta n_p)} , \\
P(\Delta n_x, \Delta n_p) &=& \int \! \int \! \int \! \int P(\Delta n_x, \Delta n_p | \alpha, \tau)P(\alpha) P(\tau)
d^2 \alpha d^2 \tau
\label{EqPnnAp}
\end{eqnarray}
yielding the conditional density matrix at the output of the first beam splitter
\begin{eqnarray}
\hat \varrho (\Delta n_x, \Delta n_p) = \int \! \int \! \int \! \int 
P(\alpha, \tau | \Delta n_x, \Delta n_p) \left|\kappa \alpha + \sqrt{1-\kappa^2}\tau \right\rangle \left\langle \kappa \alpha + \sqrt{1-\kappa^2}\tau    \right|
d^2 \alpha d^2 \tau
\label{rhoAlTau}
\end{eqnarray}

So far the results are exact. We now adopt the Gaussian approximation by assuming that the probability of photodetection for input coherent states $|\alpha \rangle$ and $|\tau \rangle$ is
\begin{eqnarray}
\label{s116}
P(\Delta n_x, \Delta n_p | \alpha, \tau) \approx \frac{1}{2\pi D^2}\exp \left[
-\frac{\left(\Delta n_x - \langle \Delta n_x\rangle  \right)^2 + \left(\Delta n_p - \langle \Delta n_p\rangle  \right)^2}{2D^2}
\right] .
\end{eqnarray}
This function depends on $\alpha$ and $\tau$ through $\langle \Delta n_x\rangle$, $\langle \Delta n_x\rangle$, as well as through the Gaussian width~$D^2$. 
The next step in the approximation is to replace in (\ref{EqD2}) the value of $(\tilde x^2 + \tilde p^2)$ by its average, assuming thermal distributions for $\alpha$ and $\tau$, with  
\begin{eqnarray}
\langle x^2 \rangle = \langle p^2 \rangle &=& \bar n, \\
\langle x^2_{\tau} \rangle = \langle p^2_{\tau} \rangle &=& \bar n_{\tau}, \\
\langle x x_{\tau} \rangle = \langle p p_{\tau} \rangle &=& 0. 
\end{eqnarray}  
We find
\begin{eqnarray}
\langle \tilde x^2\rangle = \langle \tilde p^2\rangle = (1-\kappa^2)\bar n + \kappa^2 \bar n_{\tau}.
\end{eqnarray}
The gaussian width in Eq. (\ref{s116}) is then found to be
\begin{eqnarray}
\nonumber
D^2 &\approx & 
\kappa_D^4 \bar n_{\rm H}\bar n_{\rm LO}
+ \frac{\kappa_D^2(1-\kappa_D^2)}{2}\left( \bar n_{\rm H} + 2\bar n_{\rm LO} \right) \bar n_{\rm D}
+ 2 (1-\kappa_D^2)^2\bar n_{\rm D}^2  
+ \kappa_D^2\left[1+2 (1-\kappa_D^2)\bar n_{\rm D}\right] \beta^2
 \nonumber \\
& & 
+ \frac{\kappa_D^2}{2}\left[\kappa_D^2 ( \bar n_{\rm H} +2 \bar n_{\rm LO}) + 2(1-\kappa_D^2)\bar n_{\rm D}+1 \right] 
\left[ (1-\kappa^2)\bar n + \kappa^2 \bar n_{\tau} \right]
+ \frac{\kappa_D^2}{2}\bar n_{\rm H} + \kappa_D^2\bar n_{\rm LO} + 2 (1-\kappa_D^2)\bar n_{\rm D} .
\label{EqD2p1}
\end{eqnarray}
In the case of perfect photodetectors ($\kappa_D=1$) and no fluctuations of the LO or the BS ($\bar n_{\rm LO}=\bar n_{\rm H}=0$) this expression reduces to 
\bn
D^2= \beta^2+(1-\kappa^2)\bar n/2.
\en

The mean values of quadratures $\hat x$ and $\hat p$ in state (\ref{rhoAlTau}) are
\begin{eqnarray}
\langle x \rangle &=& \int \! \int \! \int \! \int \left\langle \kappa \alpha + \sqrt{1-\kappa^2}\tau    \right|
\hat x \left|\kappa \alpha + \sqrt{1-\kappa^2}\tau \right\rangle P(\alpha, \tau | \Delta n_x, \Delta n_p) 
d^2 \alpha d^2 \tau
\nonumber \\
&=& \int \! \int \! \int \! \int 
( \kappa x + \sqrt{1-\kappa^2}x_{\tau}  ) P(x,p,x_{\tau},p_{\tau} | \Delta n_x, \Delta n_p) 
dx dp dx_{\tau} dp_{\tau} , \\
\langle p \rangle &=& \int \! \int \! \int \! \int \left\langle \kappa \alpha + \sqrt{1-\kappa^2}\tau    \right|
\hat p \left|\kappa \alpha + \sqrt{1-\kappa^2}\tau \right\rangle P(\alpha, \tau | \Delta n_x, \Delta n_p) 
d^2 \alpha d^2 \tau
\nonumber \\
&=& \int \! \int \! \int \! \int 
( \kappa p + \sqrt{1-\kappa^2}p_{\tau}  ) P(x,p,x_{\tau},p_{\tau} | \Delta n_x, \Delta n_p) 
dx dp dx_{\tau} dp_{\tau} .
\end{eqnarray}
The conditional probability in the Gaussian approximation, using (\ref{EqD2p1}), is
\begin{eqnarray}
P(x,p,x_{\tau},p_{\tau} | \Delta n_x, \Delta n_p)
&=& A \exp \left\{
-\frac{\left[\Delta n_x - \kappa_D^2 \beta \left( \sqrt{1-\kappa^2} x - \kappa x_{\tau} \right)\right]^2 + \left[\Delta n_p -  \kappa_D^2 \beta \left( \sqrt{1-\kappa^2} p - \kappa p_{\tau} \right)  \right]^2}{2D^2}
\right\} \nonumber \\
& &\times \exp \left( - \frac{x^2+p^2}{2\bar n} \right) 
\exp \left( - \frac{x_{\tau}^2+p_{\tau}^2}{2\bar n_{\tau}} \right)
\end{eqnarray}
which yields
\begin{eqnarray}
\langle x \rangle &=& \frac{\kappa_D^2 \beta \kappa \sqrt{1-\kappa^2}(\bar n-\bar n_{\tau})}{\kappa_D^4 \beta^2 [(1-\kappa^2)\bar n + \kappa^2 \bar n_{\tau}]+ D^2}\Delta n_x, \\
\langle p \rangle &=& \frac{\kappa_D^2 \beta \kappa \sqrt{1-\kappa^2}(\bar n-\bar n_{\tau})}{\kappa_D^4 \beta^2 [(1-\kappa^2)\bar n + \kappa^2 \bar n_{\tau}]+ D^2}\Delta n_p.
\end{eqnarray}
Since these expressions contain $(\bar n-\bar n_{\tau})$, one sees that no work can be obtained if $\bar n =\bar n_{\tau}$.

Integrating (\ref{EqPnnAp}) one finds
\begin{eqnarray}
P(\Delta n_x, \Delta n_p) &=&  \frac{1}{2\pi \sigma^2_{\Delta n}}
\exp \left[ -\frac{\Delta n_x^2 + \Delta n_p^2}{2  \sigma^2_{\Delta n}} \right] , \\
 \sigma^2_{\delta n}&=& \kappa_D^4 \beta^2 \left[ (1-\kappa^2)\bar n + \kappa^2 \bar n_{\tau} \right] + D^2 .
 \label{PWN}
\end{eqnarray}
The average work obtained by displacing the conditional state is
\begin{eqnarray}
W &=& \frac{1}{2}\int \! \int \left( \langle x \rangle^2 +  \langle p \rangle^2 \right)
P(\Delta n_x, \Delta n_p) d \Delta n_x d \Delta n_p- 2\beta^2 \nonumber \\
&=& \frac{\kappa_D^4 \beta^2 \kappa^2 (1-\kappa^2)(\bar n-\bar n_{\tau})^2}{\kappa_D^4 \beta^2 \left[ (1-\kappa^2)\bar n + \kappa^2 \bar n_{\tau} \right]+D^2}- 2\beta^2 .
\end{eqnarray}
This result shows the general dependence of the extractable work on final photodetection efficiency and on thermal noise in the Gaussian approximation. 
\subsubsection{Imperfect photodetection, no thermal noise}
Assume $\bar n_{\rm H} = \bar n_{\rm LO} = \bar n_{\rm D} = \bar n_{\tau} = 0$. Then Eq. (\ref{EqD2p1}) reduces to
\begin{eqnarray}
D^2 = \frac{\kappa_D^2}{2}(1-\kappa^2)\bar n + \kappa_D^2 \beta^2
\end{eqnarray}
and we obtain
\begin{eqnarray}
W &=& \kappa_D^2 \frac{2 \beta^2 \kappa^2 (1-\kappa^2)\bar n^2}{2\beta^2 + (1-\kappa^2)
(1+ 2\beta^2 \kappa_D^2)\kappa_D^2 \bar n} - 2\beta^2.
\end{eqnarray}
 For $\kappa_D <1$ the extractable work decreases because part of the energy is wasted by the imperfect photodetection.

\subsubsection{Thermal noise at the first beam splitter}
Assume $\bar n_{\rm H} = \bar n_{\rm LO} = \bar n_{\rm D} = 0$, $\kappa_D = 1$, and general $\bar n_{\tau}$. Then
\begin{eqnarray}
D^2 = \frac{1}{2}\left[(1-\kappa^2)\bar n + \kappa^2 \bar n_{\tau} \right] + \beta^2
\end{eqnarray}
and 
\begin{eqnarray}
\label{s133}
W &=&  \frac{2 \beta^2 \kappa^2 (1-\kappa^2)(\bar n - \bar n_{\tau})^2}{2\beta^2 + (1+2\beta^2)[(1-\kappa^2)\bar n + \kappa^2 \bar n_{\tau}]}- 2\beta^2 ,
\end{eqnarray}
One can see from Eq. (\ref{s133}) that the thermal imbalance between the inputs of the first BS ($\bar n \neq \bar n_{\tau}$) is essential for any work extraction. This has a clear explanation: if thermal fields of the same temperature enter the inputs of the beam splitter, independent thermal fields of the same temperature also leave the outputs. There is no correlation between these fields, and therefore  measurements of one mode cannot give us any information as to where we should displace the field of the remaining mode.

\subsubsection{Thermal noise at the homodyne BS}
Assume $\bar n_{\tau} = \bar n_{\rm LO} = \bar n_{\rm D} = 0$, $\kappa_D = 1$, and thermal noise with any $\bar n_{\rm H}$. Then
\begin{eqnarray}
D^2 = \frac{1}{2}(1-\kappa^2)\bar n (1+\bar n_{\rm H}) + \frac{\bar n_{\rm H}}{2} + \beta^2
\end{eqnarray}
and 
\begin{eqnarray}
W &=&  \frac{2 \beta^2 \kappa^2 (1-\kappa^2)\bar n ^2}{2\beta^2 + (1-\kappa^2)\bar n (1+2\beta^2 + \bar n_{\rm H}) + \bar n_{\rm H}}- 2\beta^2 .
\end{eqnarray}
One can see that presence of thermal noise  at the homodyne BS decreases the extractable work.

\subsubsection{Thermal noise in the local oscillators}
Assume $\bar n_{\tau} = \bar n_{\rm H} = \bar n_{\rm D} = 0$, 
$\kappa_D = 1$, and general $\bar n_{\rm LO}$, then
\begin{eqnarray}
D^2 = \frac{1}{2}(1-\kappa^2)\bar n (1+2\bar n_{\rm LO}) + \bar n_{\rm LO} + \beta^2
\end{eqnarray}
and 
\begin{eqnarray}
W&=&  \frac{2 \beta^2 \kappa^2 (1-\kappa^2)\bar n ^2}{2\beta^2 + (1-\kappa^2)\bar n (1+2\beta^2 + 2\bar n_{\rm LO}) + 2\bar n_{\rm LO}} - 2\beta^2.
\end{eqnarray}

\subsubsection{Dark counts in the photodetectors}
Assume 
\begin{equation}
\label{136}
 \bar n_{\tau} = \bar n_{\rm H} = \bar n_{\rm LO} = 0,
\end{equation}
 and $\kappa_D \approx 1$, with large $\bar n_{\rm D}$ such that
 \begin{equation}
 (1-\kappa_D^2)\bar n_{\rm D} \equiv \bar N_{\rm D}. 
 \end{equation}
This is a model of a detector which reacts to any input but on top of that also has dark counts. One finds then
\begin{eqnarray}
D^2 = \frac{1}{2}(1-\kappa^2)\bar n (1+2\bar N_{\rm D}) + (1+2\bar N_{\rm D}) \beta^2 + 2\bar N_{\rm D} +2\bar N_{\rm D}^2
\end{eqnarray}
and
\begin{eqnarray}
W &=&  \frac{2 \beta^2 \kappa^2 (1-\kappa^2)\bar n ^2}{2\beta^2 + (1-\kappa^2)\bar n (1+2\beta^2 + 2\bar N_{\rm D}) + 4 \bar N_{\rm D} \left( 1+ \beta^2 +\bar N_{\rm D} \right)}- 2\beta^2 .
\end{eqnarray}
The dark counts decrease the available work and their influence is even stronger than the fluctuations in the local oscillators.

\begin{figure}
\centering
  \includegraphics[width=10cm]{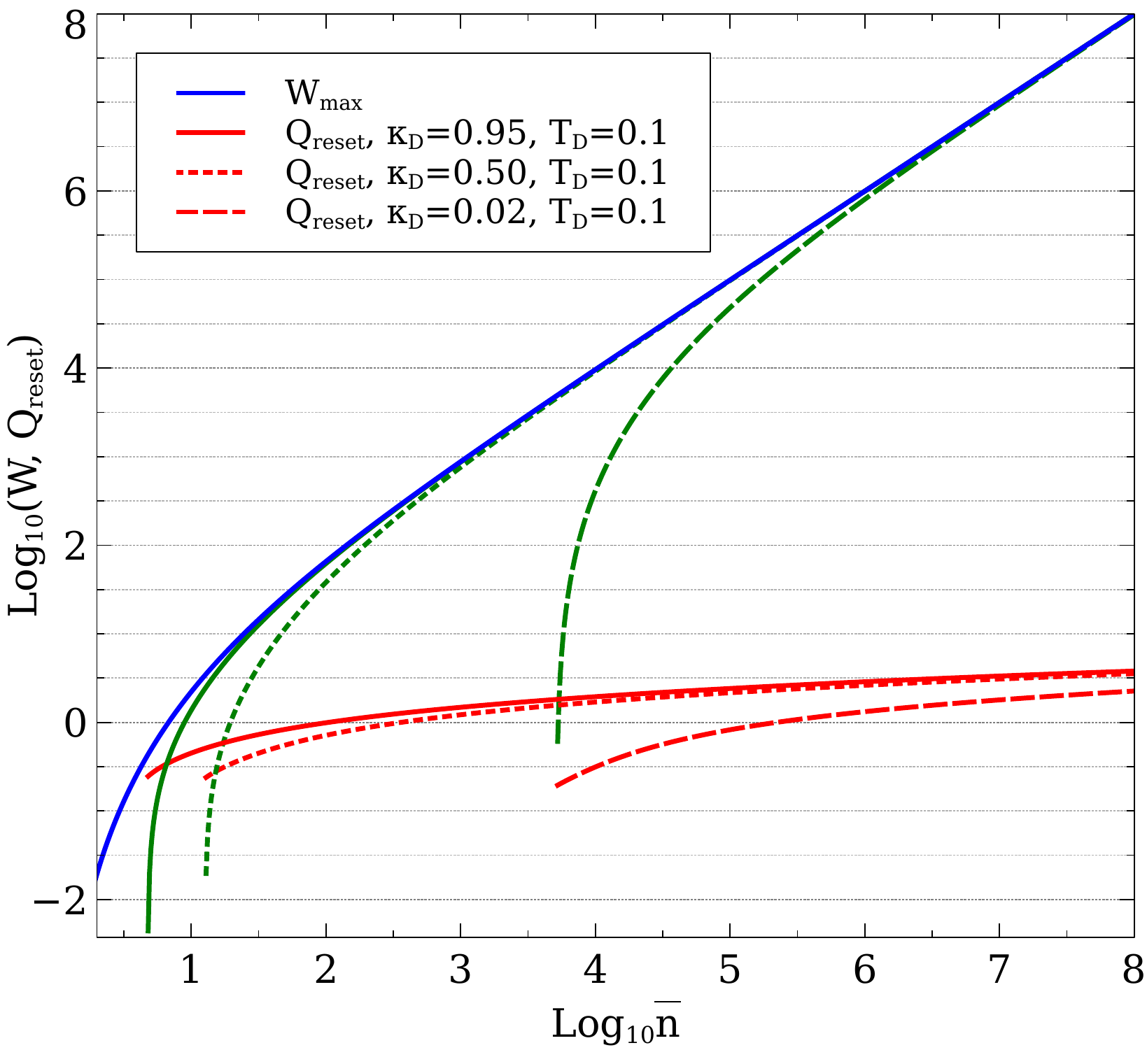}
\caption{ WOF work extraction and resetting cost (${\rm Log}_{10}$ scale) as a function of ${\rm Log}_{10} \bar n$: Blue-line $W_{max}$ (Eq. (3) of the main text). Red lines (solid, dotted and dashed) show $Q_{\rm reset}$ for different $\kappa_D$ and scaled detector temperatures $k_BT_D/\hbar \omega$. Green lines (solid, dotted and dashed) show $W_{\rm net}\equiv W-Q_{\rm reset}$ for the same parameter values as the corresponding red lines. As oppossed to the fast convergence of $W_{\rm net}$ to $W_{\rm max}$,  $Q_{\rm reset}$ does not grow much with ${\rm Log}_{10} \bar n$.}
\label{F2S}
\end{figure}
\subsection{Dissipated heat for detector resetting}
\label{S-Landauer}

The mean photon number in the detectors increases due to the incident photons by in the setup described by SM-5 the amount
\begin{eqnarray}
\Delta \bar n_d=\kappa_D^2 \left( \frac{1-\kappa^2}{4}\bar n + \frac{\kappa^2}{4}\bar n_{\tau}+\frac{1}{4}\bar n_H + \frac{1}{2}\beta^2 + \frac{1}{2}\bar n_{\rm LO} \right)
+ (1-\kappa_D^2)\bar n_{\rm D}.
\label{140}
\end{eqnarray}
In the absence of spurious or thermal noise in the detectors, as per (\ref{136}), (\ref{140}) amounts to
\begin{eqnarray}
\label{deten}
\Delta \bar n_d=\kappa_D^2 \left( \frac{1-\kappa^2}{4}\bar n + \frac{1}{2}\beta^2 \right),
\end{eqnarray}
where $\kappa_D^2$ reduces (diverts) the energy absorbed by the detectors (see Discussion in the main text).

The total entropy increase (heatup) due to detection (in bits) assuming the highest-entropy state with the mean energy $\bar n_d+\Delta \bar n_d$, is the mean information stored in the detectors
\begin{eqnarray}
I=\frac{\Delta S}{\ln 2} \approx 4 \frac{S(\Delta \bar n_d+\bar n_d)-S(\bar n_d)}{ \ln 2},
\end{eqnarray}
where the initial entropy of the each detector prior to the measurement is 
\begin{equation}
S(\bar n_d)=(\bar n_d +1)\ln (\bar n_d +1) - \bar n_d \ln \bar n_d.
\end{equation} 
According to Landauer's principle 
the minimum amount of heat $Q_{\rm reset}$ dissipated to the the environement at temperature $T_D$ when resetting the detectors is $k_BT_D \ln 2$ per bit. In this way we get the Landauer resetting energy cost (Fig. \ref{F2S})
\begin{eqnarray}
 Q_{\rm reset}= I k_BT_D \ln 2.  
\end{eqnarray}
Using (\ref{Apxi}) and (\ref{EOpteta}) in the large $\bar n$ limit and for $\kappa_D=1$, from Eq. (\ref{deten}) one gets 
\begin{equation}
 \Delta \bar n_d=\frac{\sqrt{\bar n}}{2}.
\end{equation}
In the large $\bar n$ limit, using $\Delta S \approx 4 S(\Delta \bar n_d)$ and
using $S(\Delta \bar n)\approx1 + \ln \Delta \bar n$, one finds that the mean information stored (the total entropy change) in the four detectors is then
\begin{equation}
I=\frac{\Delta S}{\ln 2} \approx \frac{1}{2} \ln \frac{\bar n}{4}.
\end{equation}

The state-of-the-art  on the  resetting energy cost may  be estimated from refs. \cite{Natarajan, Wolff2020} : Detection of a  single photon destroys the current in the superconducting nanowire, which must be restored by resetting.  The current is  ca.  3 microamps,  passing though a circuit of inductance $L = 500$ nH, and  has  an  energy of  $\frac{1}{2} LI^2\sim 2\times10^{-18}$ J.  This  resetting  energy  is an  order of magnitude higher than the energy of the detected photon. The corresponding  resetting time is of a few ns.

\subsection{Comparison of WOF with Szilard/ Maxwell-Demon engine based on
thermal-noise photodetection}
In  an experimentally tested Szilard/ Maxwell Demon engine based on photodetection \cite{Vid2016}, two thermal beams (fields) are used as input, each having on average $\bar n$ photons. A single-photon  click or no-click is registered for each beam (two bits of information).
  In  the simplest case, the click  probability is $1/2$. In this case, if the respective  detector clicks then $\bar n$ of the corresponding output  field increases to $(3/2) \bar n$, and if it does not, it decreases to $(1/2) \bar n$. Thus, if one detector clicks and the other one does not (in $50\%$ of the cases), there is an average difference $\bar n [ = (3/2-1/2) \bar n]$ leading to a  net photocurrent that charges a capacitor. If both or neither of the  detectors click, there
is no difference in the average  output  fields  and no mean  
photocurrent. Thus, the energy convertible to photocurrent  is  $(1/2) \bar n$.  Since the  two beams have in  total $2 \bar n$, only  $1/4$ of the input energy is exploited for work, i.e. the efficiency bound is $1/4$. Optimization of the click probabilities is achievable in an  arrangement where one detector fires with probability $1/3$ and the other with probability $2/3$ (thus instead of collecting two bits of information just $\sim 1.8$ bits are used) and  on average
$(16/27) \bar n$ quanta are converted into work, thus yielding  the  efficiency  bound  of such a machine to  $\sim 0.3$,  but nowhere near the WOF efficiency bound in Eq. (5) in the main text.


Among other factors, the efficiency of this machine is lowered by the fundamental Shockley-Queisser bound on photocurrent  efficiency \cite{shockley}. 
By contrast, WOF is  much less susceptible to efficiency reduction due to this bound: WOF
only converts a small fraction of the photonic  input into a photocurrent and extracts  the work   in the form of  output  light (SM- 3), as opposed to the Szilard/Maxwell-Demon
machine that converts the photonic  output into electric energy.

\subsection{Comparison of wof  with otto Heat engines}

Here we compare the performance bounds of WOF with finite-time Otto cycles that can bridge reciprocal and continuous HE models \cite{Rezek}. For the comparison, we set (see Eq. (5) and Discussion of the main text) $E_{\rm in}=k_B T_h$, $E_{\rm rem}=k_B T_c$, so that 
\begin{equation}
\label{147}
 \eta_{\rm Carnot}=\eta_{\rm reverse}=1-\frac{E_{\rm rem}}{E_{\rm in}}>\eta_{\rm max}.
\end{equation}
It is customary to distinguish between two extreme regimes of the Otto HE:

1){\it Frictionless regime (FL)}: In this regime, the optimal time duration of the two adiabatic strokes is ($O(1/\sqrt{\omega_h \omega_c})$, where $\omega_h, \omega_c$ are the working medium (WM) frequency values after and before compression. 
In the high temperature limit, when $\hbar\omega/k_BT_{h(c)}\ll1$, the work production is optimized at the compression ratio $\mathcal{C}=\frac{\omega_h}{\omega_c}=\sqrt{\frac{T_h}{T_c}}$, leading to the efficiency 
\begin{equation}
\label{148}
 \eta_{\rm Otto}^{(\rm FL)}=1-\sqrt{\frac{T_c}{T_h}},
\end{equation}
which is the efficiency at maximum power according to Novikov \cite{novi} and Curzon-Ahlborn \cite{curzon}, and can be well below $\eta_{\rm Carnot}$.

For sufficiently large relaxation rates $\Gamma_h$, $\Gamma_c$ to the hot and cold baths, the optimal power condition is obtained for the bang-bang solution \cite{Feldmann} where vanishingly small time is allocated to the isochores. The work extraction per cycle is then 
\begin{equation}
 W_{\rm Otto}^{(\rm FL)}=G_W\frac{\Gamma_h\Gamma_c}{(\sqrt{\Gamma_h}+\sqrt{\Gamma_c})^2}.\tau_{cyc},
\end{equation}
where $G_W=\frac{\hbar}{2}(\omega_h-\omega_c)({\rm coth}(\frac{\hbar \omega_h}{2 k_B T_h})-{\rm coth}(\frac{\hbar \omega_c}{2 k_B T_c}))$, and $\tau_{cyc}=\tau_{iso}+\tau_{adi}\approx \tau_{adi}$ is the total cycle time, $\tau_{iso}$ and $\tau_{adi}$ being the isochoric and adiabatic stroke times  respectively.
The factor $\frac{\Gamma_h\Gamma_c}{(\sqrt{\Gamma_h}+\sqrt{\Gamma_c})^2}$ is due to short time allocation to the isochores which reduces the maximum work output $G_W$ corresponding to the extremely slow (quasistatic) cycle.
The maximum power for $\Gamma_h=\Gamma_c\equiv \Gamma$ is then given by
\begin{equation}
 P_{max}^{(\rm FL)}=G_W\frac{\Gamma}{4}.
\end{equation}
 For high tempeartures, $P_{max}$ is upper bounded by
\begin{equation}
 P_{max}^{(\rm FL)}\lesssim k_B T_h\frac{\Gamma}{4},
\end{equation}
approaching the equality for high compression ratio, i.e., for $T_h\gg T_c$.

If we identify the WM relaxation time $1/\Gamma$ with $\tau_{\rm reset}$ of the detectors in WOF \cite{Gaudenzi2018}, then the Otto-cycle work output at maximal power is {\it $4$
times smaller} than either the maximal (quasistatic) Otto work output $G_W =k_B T_h$ or the WOF maximal work output $W_{\rm max} \simeq k_B T_h$ at high $T_h$  within the same time window $1/\Gamma$.

2){\it Sudden regime (S)}: In this regime the adiabatic strokes are performed with almost zero time allocation and therefore work production is reduced due to friction. At high temperatures the optimal compression ratio for the maximum work produced is achieved for $\mathcal{C}=(\frac{T_h}{T_c})^{1/4}$, leading to the efficiency that can be much inferior to both (\ref{147}) and (\ref{148}):
\begin{equation}
 \eta_{\rm Otto}^{(\rm S)}=\frac{1-\sqrt{\frac{T_c}{T_h}}}{2+\sqrt{\frac{T_c}{T_h}}} \leq1/2.
\end{equation}
 Taking the frictional cost into account, the optimized work per cycle is given by (again for $\Gamma=\Gamma_h=\Gamma_c$)
\begin{equation}
 W_{\rm Otto}^{(\rm S)}\approx k_B T_c \frac{\Gamma \tau_{cyc}}{2}\frac{\mathcal{C}^2-1}{2\mathcal{C}^2}(\frac{T_h}{T_c}-\mathcal{C}^2).
\end{equation}
For high compression ratio this expression is again bounded by
\begin{equation}
 W_{\rm Otto}^{(\rm S)}\lesssim k_B T_h \frac{\Gamma \tau_{cyc}}{4}.
\end{equation}
The corresponding upper bound of the power that is achieved with vanishing cycle time, such that the Otto cycle reaches the limit of {\it continuous operation}, is
\begin{equation}
 P_{\rm max}^{(\rm S)}\lesssim k_B T_h \frac{\Gamma}{4}.
\end{equation}
Thus, the Otto engine yields also in this regime at most $\sim 1/4$ of the power delivered by WOF at high $T_h$, although there is great mismatch of the time scales:
\begin{equation}
 \tau_{reset}=1/\Gamma\gg \tau_{cyc}\rightarrow 0.
\end{equation}
\end{widetext}

\end{document}